# Reservoir model for two-dimensional electron gases in quantizing magnetic fields: a review

**W. Zawadzki[1], A. Raymond[2], and M. Kubisa[*,3]**

[1] Institute of Physics, Polish Academy of Sciences, 02668 Warsaw, Poland
[2] Universite Montpellier 2, Laboratoire Charles Coulomb UMR5221, F-34095, Montpellier, France
[3] Institute of Physics, Wroclaw University of Technology, 50-370 Wroclaw, Poland

* Corresponding author: e-mail Maciej.Kubisa@pwr.wroc.pl

We collect and review works which treat two-dimensional electron gases in quantum wells (mostly GaAs/GaAlAs heterostructures) in the presence of quantizing magnetic fields as open systems in contact with outside reservoirs. If a reservoir is sufficiently large, it pins the Fermi level to a certain energy. As a result, in a varying external magnetic field, the thermodynamic equilibrium will force oscillations of the electron density in and out of the quantum well (QW). This leads to a number of physical phenomena in magneto-transport, interband and intraband magneto-optics, magnetization, magneto-plasma dispersion, etc. In particular, as first proposed by Baraff and Tsui, the density oscillations in and out of QW lead to plateaus in the Integer Quantum Hall Effect (IQHE) at values observed in experiments. The gathered evidence, especially from magneto-optical investigations, allows us to conclude that, indeed, in most GaAs/GaAlAs hetrostructures one deals with open systems in which the electron density in QWs oscillates as the magnetic field varies. Relation of the density oscillations to other factors, such as electron localization, and their combined influence on the quantum transport in 2D electron gases, is discussed. In particular, a validity of the classical formula for the Hall resistivity $\rho_{xy} = B/Nec$ is considered. It is concluded that the density oscillations are not sufficient to be regarded as the only source of plateaus in IQHE, although such claims have been sometimes made in the past and present. Still, our general conclusion is that the reservoir approach should be included in various descriptions of 2D electron gases in the present of a magnetic field. An attempt has been made to quote all the relevant literature on the subject.

**1 Introduction and short history** When Klaus von Klitzing discovered the Quantum Hall Effect (QHE) [1], there appeared a natural need to explain exact values of the observed quantized Hall plateaus. Among various preliminary explanations that followed, Baraff and Tsui [2] put forward a model that later, with various modifications, existed in the literature under the names of "reservoir hypothesis" or "reservoir model". Baraff and Tsui described in a self-consistent way a heterostructure of GaAs/GaAlAs type, selectively doped with donors in the GaAlAs barrier and subjected to an external magnetic field $B$ parallel to the growth direction. Their calculation showed that the ground energy of the dopant donors fixes the Fermi level in the structure, so that, when $B$ is increased and the density of states due to the Landau levels is a sequence of strong maxima, the electrons tunnel back and forth between the donors in GaAlAs barrier and the 2D electron gas in the GaAs quantum well. This means that the density $N$ of 2D electron gas oscillates and, in the field regions where $N$ grows linearly with $B$, the Hall resistance exhibits plateaus having exactly the values measured experimentally. The last feature is a simple result of the Landau level degeneracy. The result of Baraff and Tsui was confirmed within the same model by a somewhat simpler calculation of Bok and Combescot [3].

However, at the same time a different line of thought prevailed, explaining the quantum Hall phenomenon by a localization of electron states within the Landau levels, see the reviews [4, 5]. The interpretation based on the electron localization became so dominant that during a certain period it was difficult to publish different points of view, cf. Refs. [6, 7]. Still, the reservoir model has kept appearing in the literature under different names in order to explain various observations on the 2D electron gases: quantum transport [8-10], Fermi energy behaviour [11], cyclotron resonance [12, 13], interband photo-magneto-luminescence [14], magnetic susceptibilities [15], magneto-plasmon dispersion [16, 17], etc. In his well known book, Mahan [18] treats the localization and reservoir interpretations of QHE on equal footing. Recently, the electron reservoir made a convincing reappearance in monolayer graphene [19]. Thus it seems that now, when the smoke of battles over the quantum transport in 2D electron gases is not as thick as it used to be, it is a good time to write a review on the subject.

The purpose of our review is to collect and briefly discuss publications suggesting the presence of electron reservoir in various experiments on 2D electron gases. An important place is reserved to the quantum magneto-transport effects which started the whole discussion, but other phenomena are also presented. In fact, the latter are often more convincing because the charge transport is difficult to describe. It is hoped that our review will stimulate additional investigations to clarify obscure points concerning this important problem. An effort has been made to quote all the relevant literature on the subject.

**2 Constant electron density versus constant Fermi level** In the following section we consider briefly thermodynamic properties of 2D electron gases (2DEG) in two limiting situations. The first is the standard case of a constant electron density in the quantum well: $N$ = const. The second describes the case of a 2DEG in contact with an external reservoir that can "pin" the Fermi level $E_F$. In order to emphasize the main features and make calculations easier we consider an extreme case of a large reservoir that can completely fix the value of $E_F$ = const. We contrast the two situations in order to make the following considerations understandable. Finally, we quote very briefly results for a self-consistent calculation. This is done for historic reasons, since a similar calculation was performed by Baraff and Tsui, and also because it represents a realistic case realized in GaAs/GaAlAs hetrostructures.

**2.1 Constant electron density.** We consider 2DEG of non-interacting electrons in a parabolic, spherical energy band at a finite temperature $T$ in the presence of a quantizing magnetic field $B$ parallel to the growth direction. The spin degeneracy is included but it is assumed that the spin-splitting factor $g^* = 0$. Quantum wells and superlattices based on GaAs satisfy quite well these assumptions if the exchange enhancement of the $g$ value is neglected (see below). An incorporation of the spin splitting is straightforward. We assume further that only one electric subband is populated. The description is based on the work of Zawadzki and Lassnig [20]. The energetic density of states (DOS) is taken in the form of a sum of Gaussian peaks

$$\rho(E) = \frac{1}{2\pi L^2}\sum_{n,s}\sqrt{\frac{2}{\pi}}\frac{1}{\Gamma}\exp\left[-2\left(\frac{E-\lambda_{ns}}{\Gamma}\right)^2\right], \qquad (1)$$

where $L^2 = c\hbar/eB$, $\lambda_{ns} = \hbar\omega_c(n + 1/2) + (1/2)\mu_B g^* s$, $\omega_c = eB/m^*c$ is the cyclotron frequency, $n$ and $s = \pm 1$ are the Landau and spin quantum numbers, respectively, and $\Gamma$ is the broadening parameter assumed constant. Two features should be emphasized at this point. First, in addition to the Gaussian peak of DOS at each Landau level (LL), there is a common factor $B$ in front of total DOS. This means that, as $B$ increases, each LL can contain more and more electrons. Second, according to the form assumed in Eq. (1), there is no DOS between LLs if their separation $\hbar\omega_c$ is distinctly larger than $\Gamma$. This situation is illustrated in Fig. 1.

The electron density in $cm^{-2}$ is

$$N = \frac{1}{2\pi L^2}\sum_{n,s}\sqrt{\frac{2}{\pi}}\frac{1}{\gamma}\int_0^\infty \frac{\exp(-2y_{ns}^2)}{1+\exp(z-\eta)}dz, \qquad (2)$$

where $y_{ns} = (z - \theta_{ns})/\gamma$, $z = E/kT$, $\eta = E_F/kT$, $\theta_{ns} = \lambda_{ns}/kT$, $\gamma = \Gamma/kT$ are the reduced quantities. The filling factor of the system is defined as $\nu = 2\pi L^2 N$, denoting the number of occupied Landau levels (LLs). The condition of a constant electron density $N$ in the quantum well leads to an integral equation for the Fermi energy $E_F(B)$. Figure 2(a) shows the function $E_F(B)$ calculated for $m^* = 0.0665\ m_0$, $N_0 = 8 \times 10^{11}$ $cm^{-2}$, $\Gamma$ = 0.5 meV and $T$ = 6 K. It can be seen that the Fermi energy at a constant electron density oscillates quite strongly as a function of $B$.

To understand the mechanism of these oscillations let us assume that the Fermi level $E_F$ is within $n$th LL. As $B$ increases, there is more states in LLs below $E_F$ and, in consequence, the filled electron states occupy a smaller fraction of the level $n$. In fact, one can see in Fig. 2(a) that $E_F$ moves to the lower part of the level $n$. One can also say that in this region the Fermi level is "pinned" to the LL because DOS related to the latter is quite high. For a sufficiently large field $B$ all $N$ electrons can be accommodated by ($n$ - 1) levels. At this field $E_F$ falls abruptly to the ($n$ - 1) level. Since we assumed no DOS between LLs, this sharp drop of $E_F$ is vertical at low temperatures. As the field in-creases further, the process is repeated with the ($n$ - 1) level. We emphasize (see also below) that, if there were some nonvanishing DOS between the levels, the drop of $E_F$ from the $n$th to the ($n$ - 1) level would not be vertical. At higher temperatures, the drop of $E_F$ is not vertical even with no DOS between the levels.

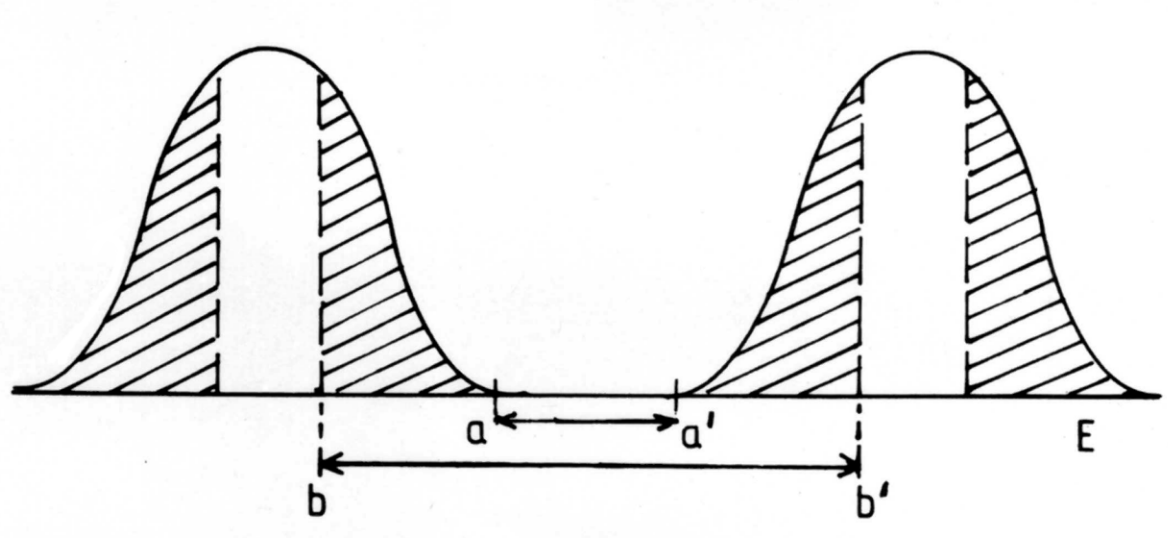

**Figure 1** Density of states for 2D electrons in a magnetic field. The shaded areas indicate localized regions. It is assumed that between the Landau levels (a-a' region) DOS is practically zero. After Ref. [7].

The free energy of the system is

$$F = NE_F - kT\int_{-\infty}^{\infty}\rho(E)\ln\left[1+\exp\left(\frac{E-E_F}{kT}\right)\right]dE. \qquad (3)$$

The magnetization of the system is $M$ = -d$F$/d$B$. One obtains

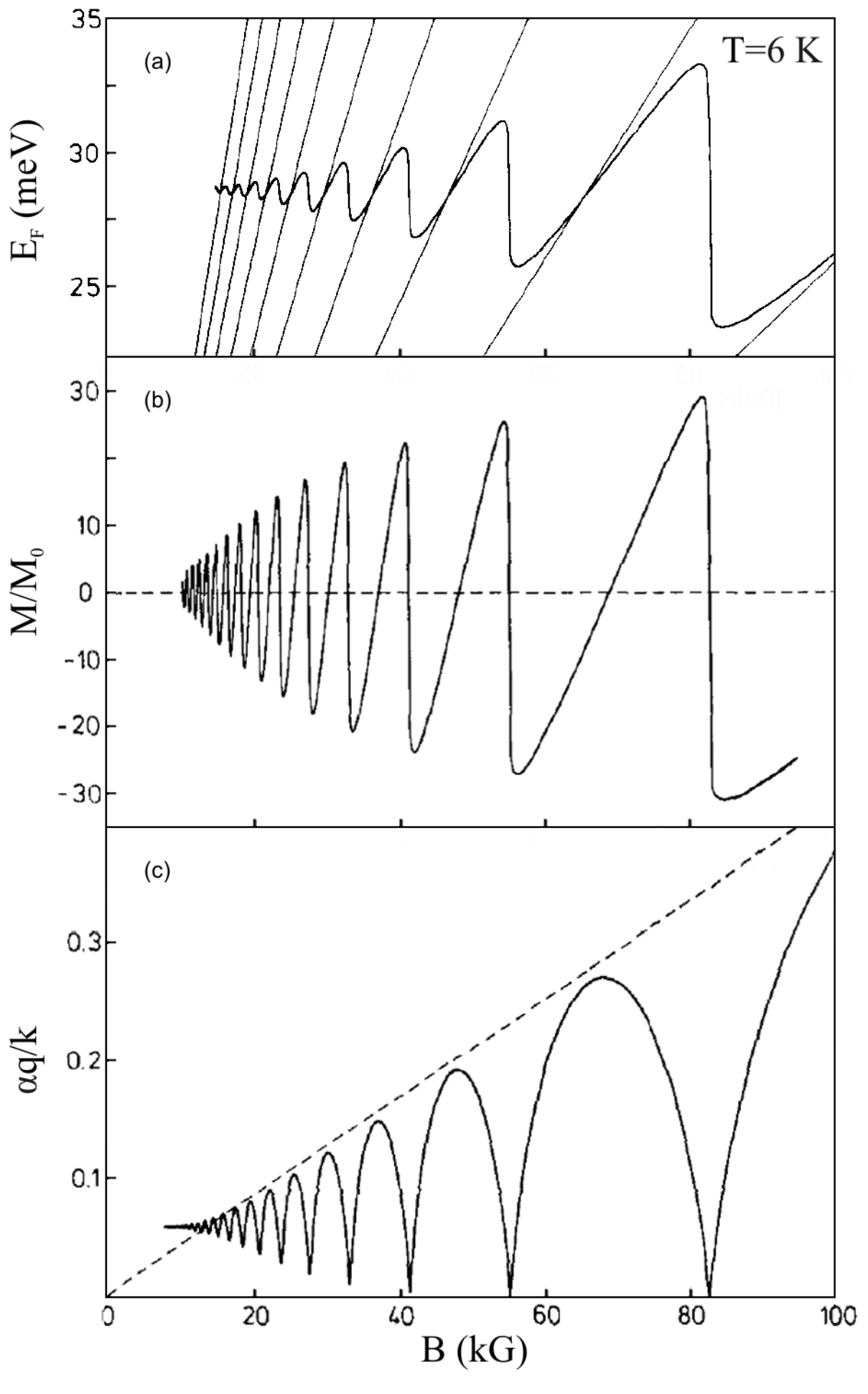

**Figure 2** (a) The Fermi energy versus magnetic field, calculated for 2DEG in GaAs at a constant electron density $N$ and $T$ = 6 K. The Landau levels are indicated. (b) Normalized magnetization of 2DEG (diamagnetic part) versus magnetic field, calculated for the same conditions and $T$ = 4.2 K. $M_0 = ekT/hc$. (c) Dimensionless thermoelectric power of 2DEG in GaAs versus magnetic field, calculated for the same conditions as in (a). The dashed line indicates maxima values of $(-e/k)(\ln 2)/\nu$, where $\nu$ is the filling factor. After Ref. [20].

$$M = \frac{ekT}{2\pi\hbar c}\sum_{n,s}\sqrt{\frac{2}{\pi}}\frac{1}{\gamma}\times \int_0^{\infty}\ln\left(1+e^{\eta-z}\right)\exp\left(-2y_{ns}^2\right)\left(1+\frac{4\theta_{nn}y_{nn}}{\gamma}\right)dz. \qquad (4)$$

Figure 2(b) shows the magnetization calculated according to Eq. (4) for the above $m^*$, $N_0$, $\Gamma$ = 0.5 meV and $T$ = 4.2 K. It can be seen that the diamagnetism of 2DEG oscillates symmetrically around the zero value. The inclusion of the spin splitting does not change this picture, it simply doubles the number of peaks. As follows from Figs. 2(a) and 2(b), the magnetization oscillations follow quite closely those of the $E_F$ - $E_0$ energy difference.

Next we want to calculate the thermoelectric power of 2DEG at high magnetic fields. It may appear surprising that a transport effect which in principle is related to carrier's scattering can be expressed by the equilibrium thermodynamic functions. This is possible because at high fields, for which $\omega_c\tau$ >> 1, one may neglect diagonal components of the conductivity tensor while the nondiagonal components do not depend on scattering. In order to calculate $\alpha(B)$ in the presence of a temperature gradient one should also include the magnetization, as showed by Obraztsov [21], see also Ref. [20]. All in all, one obtains the thermoelectric power $\alpha(B)$ at high fields in the simple form

$$\alpha(B) = -\frac{S}{eN}, \qquad (5)$$

where $N$ is the electron density given by Eq. (2) and $S$ = -(d$F$/d$T$) is the entropy of 2DEG given by

$$S = \frac{eBk}{\hbar}\sum_{n,s}\sqrt{\frac{2}{\pi}}\frac{1}{\gamma}\times \int_0^{\infty}\left[\ln\left(1+e^{\eta-z}\right)+\frac{z-\eta}{1+e^{z-\eta}}\right]\exp\left(-2y_{ns}^2\right)dz. \qquad (6)$$

Thus $\alpha$ can be readily calculated in the no-scattering limit once the Fermi energy is determined as above. Figure 2(c) shows the thermoelectric power of 2DEG (in dimensionless units) in a strong magnetic field calculated for the above parameters and the temperature $T$ = 6 K. One can deduce from Eq. (6) that the completely filled LLs (for $z$ - $\eta$ << 0) give vanishing contribution to the entropy. It is for this reason that in Fig. 2(c) the thermoelectric power (or the entropy) reaches the zero values as the Fermi energy jumps between LLs. Physically this means that the intra-level thermal excitations vanish because the levels are completely filled and the inter-level thermal excitations vanish because $kT$ is much smaller than $\hbar\omega_c$. At lower fields, when this inequality is no longer fulfilled, $\alpha(B)$ (or the entropy) does not reach zero values because of the nonvanishing inter-level excitations.

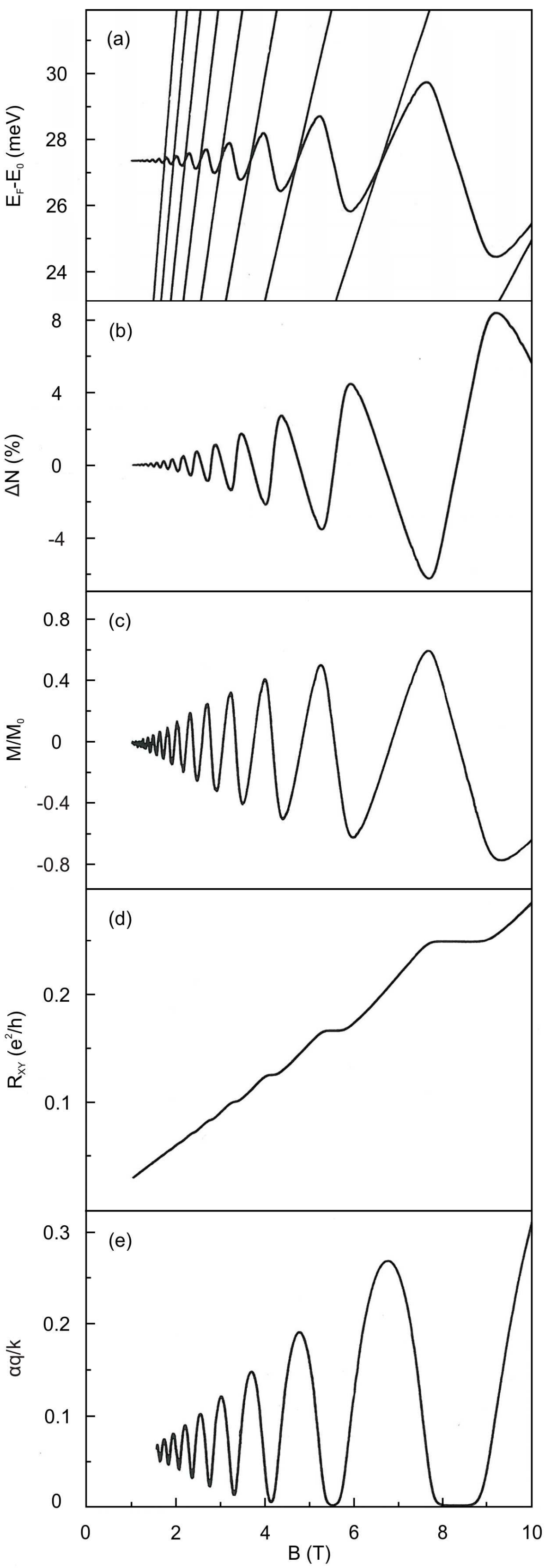

**Figure 3** Thermodynamic and magneto-transport characteristic versus magnetic field, calculated for a 2D electron gas in GaAs at $T$ = 6 K assuming that the Fermi energy is completely pinned by a large reservoir. (a) Difference between the Fermi energy $E_F$ and the bottom of the lowest electric subband $E_0$. The Landau levels are also indicated. (b) Change of electron density $N$. (c) Magnetization. (d) Quantized Hall resistance, calculated using the relation $\rho_{xy} = B/Nec$. (e) Normalized thermoelectric power. After Ref. [7].

In the above considered case of a constant 2D electron density the plateaus of QHE and the zeros of the Shubnikov-de Haas (ShdH) effect, as well as those of the thermoelectric power, are attributed to the localization regions of DOS. According to this standard theory, when the Fermi level traverses the localized region, the diagonal transport coefficients vanish while $\rho_{xy}$ has very well defined plateaus.

**2.2 Constant Fermi energy** Now we consider the opposite case of 2DEG in an open system in which a quantum well is in contact with an outside reservoir. To make things simpler and reach the main conclusions we assume that the reservoir is very large and has a well defined energy which completely pins the Fermi level at this energy. It was shown above that, when the density $N$ remains constant, the Fermi level $E_F$ oscillates as the field $B$ increases, see Fig. 2(a). It is then clear that, in order to have the Fermi level constant with the changing field, the density $N$ must oscillate. Qualitatively, the model works as follows. The oscillating electron density $N$ in QW determines the electrical potential of this well. The change of the potential results in changing the subband energy $E_0$, so that the energy interval between $E_0$ and the fixed $E_F$ changes periodically, similarly (but not identically!) to the case of constant $N$. The essential difference compared to the previously considered case is that, at a constant $N$, the Fermi level jumps between LLs whereas, at the constant Fermi level, the latter may shift more slowly between LLs. The reason is that, as $B$ increases, the electron density also increases. It will be seen below that this is the very reason for the plateaus of the Quantum Hall Effect.

A description of the reservoir approach requires a self-consistent calculation because the charge density determines the potential and the latter determines the charge transfer, i.e. the density. However, again, we use a simplified model to reach main conclusions without complicated calculations. Thus, we do not assume anything specific about the reservoir but take the Fermi level $E_F$ pinned at a constant energy from the bottom of the well. First, the subband energy $E_0$ is calculated for the initial density $N_0$ at $B$ = 0 using the variational trial wave function proposed by Ando [22]. Next, the Fermi energy is evaluated as $E_F = E_0 + N_0/D_0$, where $D_0 = m^*/(\pi\hbar^2)$ is DOS at $B$ = 0. This value of $E_F$ is assumed to remain constant in all subsequent calculations. Since the magnetic field modifies DOS, the energy difference $E_F$ - $E_0$, the energy $E_0$ and the electron density $N$ will change with $B$ . For a given $B \neq 0$, one calculates the energy $E_0$ for an input density $N_1$ and then counts the density $N_2$ filling the Landau levels between $E_0$ and $E_F$.

The value $N_1$ is then changed until $N_1 = N_2 = N$ ($B$). The potential of the well, required to calculate the subband energy, is determined by three parameters: density $N$, the offset energy $V_0$ at the GaAs /GaAlAs interface and a depletion charge $N_{depl}$. The used values are $V_0$ = 257 meV, and $N_{depl} = 6 \times 10^{10}$ cm$^{-2}$. Other parameters are the same as those given above to facilitate a comparison with the previous case. The following results are quoted after Ref. [7].

Figure 3(a) shows the calculated difference between the Fermi energy $E_F$ and the bottom of electric subband $E_0$ versus magnetic field $B$ for $T$ = 6 K. It can be seen that, in contrast to the situation shown in Fig. 2(a), this energy difference does not "jump" vertically between LLs on the higher energy side, although it is still assumed that DOS between LLs vanishes. The reason for the relatively slow decrease of $E_F - E_0$ can be understood from Fig. 3(b), which shows the calculated corresponding 2D electron density $N$ for the same scale of magnetic fields. It is seen that, as $E_F - E_0$ decreases with the field, the density $N$ in the well grows linearly with the field. Looking at Fig. 3(a) one should realize that, as the field $B$ increases and the given LL "arrives" near the constant value $E_F$, the electrons go to the reservoir and the subband energy $E_0$ begins to move down in such a way that the LL energy $E_n$ in the absolute energy scale remains almost horizontal, so that $E_n$ is "pinned" to $E_F$. This feature is a consequence of the large peak-like DOS near the energy $E_n$, as explicitly shown by Popov [23].

In Fig. 3(c) we show the corresponding magnetization calculated for the same conditions. It can be seen that, similarly to the dependences shown in Figs. 2(a) and 2(b), the behaviour of magnetization closely follows that of the difference $E_F - E_0$. The important point is that the de Haas-van Alphen (dHvA) oscillations in the two regimes have distinctly different slopes on the high-field sides.

Figure 3(d) shows the calculated ratio of $B/Nec = \rho_{xy}$ which, in the standard classical theory of magneto-transport, gives the off-diagonal component of resistance tensor describing the Hall effect. It is seen that the ratio of $B/Nec$, plotted as a function of the field $B$, exhibits plateaus. The origin of the plateaus is seen in Fig. 3(b): when $N$ increases linearly with $B$ the ratio $B/Nec$ is a constant. As mentioned above, $N$ grows linearly with $B$ when $E_F$ is between LLs. Since one LL contains $2Be/hc$ electrons (including the spin degeneracy), when there are $i$ LLs below $E_F$, the number of electrons is exactly $2Bei/hc$, so that $\rho_{xy} = B/Nec = h/2e^2 i$. These are the measured plateau values of QHE. A subtle and non-obvious point in the above reasoning is, that it assumes $\rho_{xy}$ to measure the electron density $N$ at *all* values of $B$ whereas, according to the standard interpretation, the Hall effect does *not* measure $N$ in the field region of a plateau. We come back to this point later.

Finally, Fig. 3(e) shows the thermoelectric power $\alpha(B)$ of 2DEG versus magnetic field intensity, calculated according to Eq. (5). It is seen that $\alpha(B)$ vanishes when $\rho_{xy}$ goes through plateaus. The reason is that the entropy $S$, which mostly determines $\alpha(B)$ at high fields, vanishes when the Fermi level $E_F$ is between LLs where the density of states is assumed to be zero. This behaviour is in contrast to the behaviour shown in Fig. 2(c) for the constant density $N$, where $\alpha(B)$ only touches the zero values.

We want to mention here a point which is of importance for various interpretations. If there exists a nonvanishing density of localized states between the Landau levels (as assumed in many investigations), it will also "slow down" the drop of $E_F$ between LLs as the field $B$ increases. Thus non-vanishing DOS between LLs results in the behaviour of some effects similar to that produced by the electron transfer from a reservoir. This is especially pronounced in the behaviour of magnetization as a function of $B$, see below.

**2.3 Self-consistent approach** Finally, we briefly quote results of the self-consistent approach in the presence of a magnetic field. As mentioned in the Introduction, the latter was initiated by Baraff and Tsui [2]. A similar treatment (somewhat simplified) was given by Bok and Combescot [3] and a more complete one by Xu [24]. Here we quote the results of Sabin del Valle and de Dios-Leyva [25] describing GaAs/GaAlAs heterojunctions which assumes neither constant electron density $N$ nor the constant Fermi energy $E_F$ as the field changes. We quote this work because it shows explicitly the oscillating 2D electron density in the well for different values of the spacer separating donors in the barrier from the GaAs/GaAlAs interface. A self-consistent calculation is of interest for two important reasons. First, it represents an intermediate case between the two extreme and idealized situations considered above. Second, it represents a specific open system in which QW is in contact with the realistic reservoir.

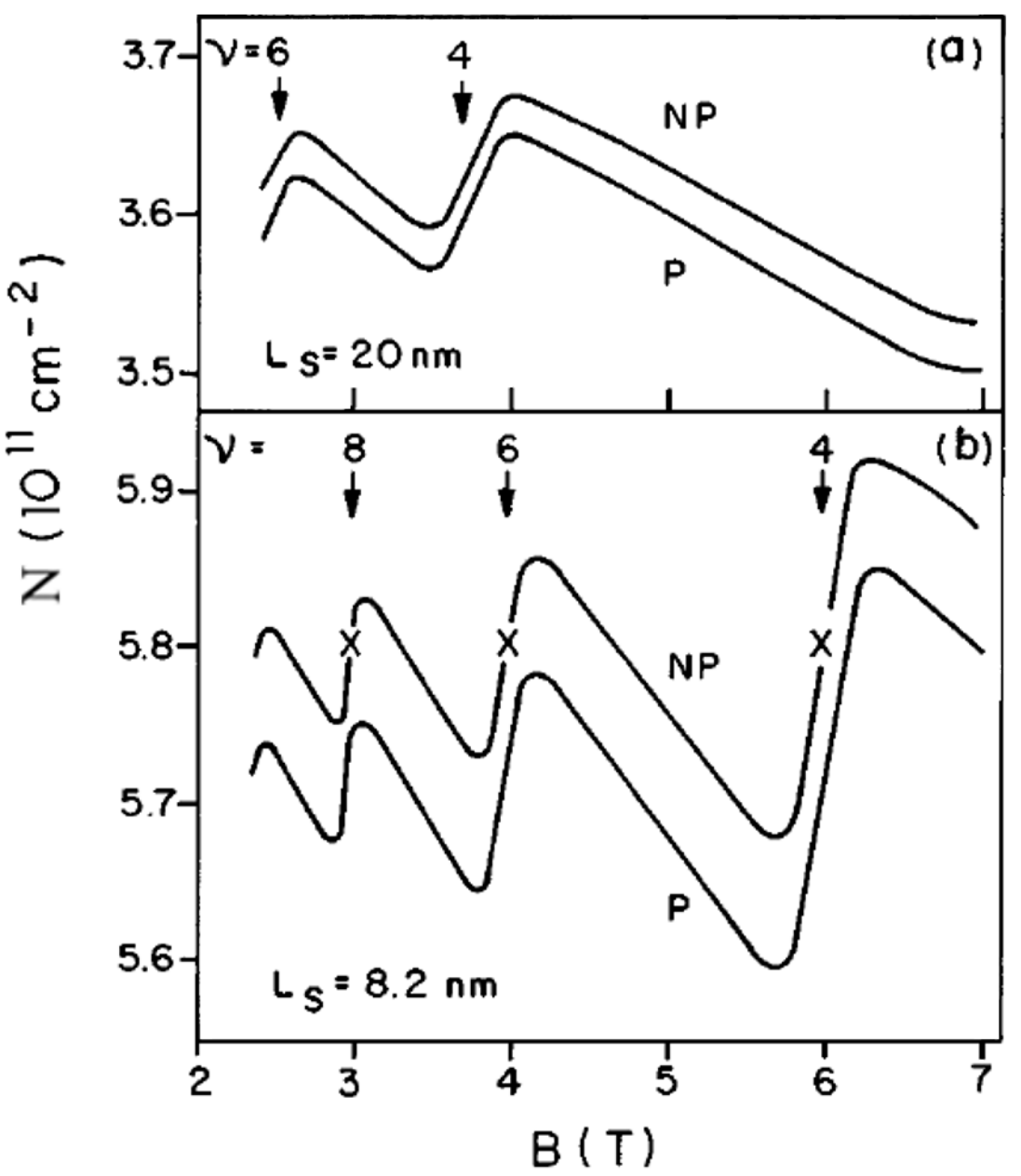

**Figure 4** Magnetic-field dependence of the electron density $N$ in GaAs/GaAlAs heterojunction, calculated with the use of parabolic (P curves) and nonparabolic (NP curves) models for the conduction band of GaAs. Spacer length: (a) $L_S$ = 20 nm and (b) $L_S$ = 8.2 nm. Experimental results for $N$ obtained from ShdH measurements [26] are marked by crosses. The calculated filling factors $\nu$ are indicated. After Ref. [25].

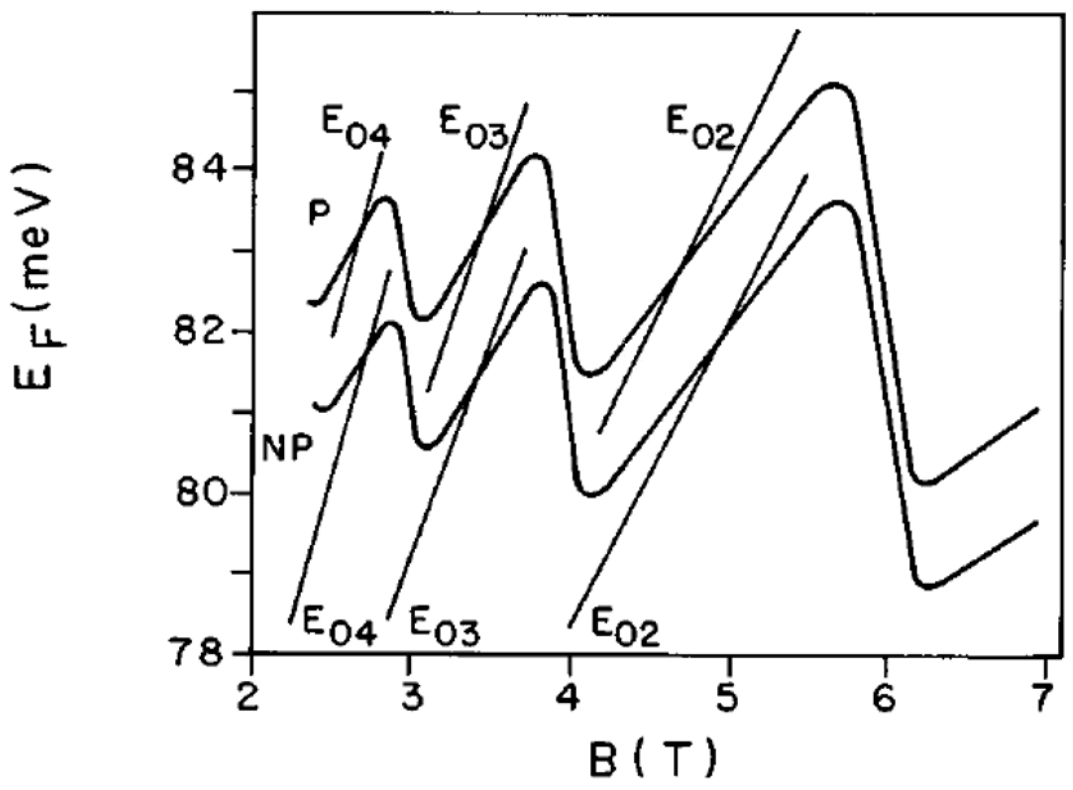

**Figure 5** Magnetic-field dependence of the Fermi energy and Landau energies for the spacer length $L_S$ = 8.2 nm, calculated with the use of parabolic (P curves) and nonparabolic (NP curves) band models for GaAs. After Ref. [25].

In a realistic situation both the 2D density $N$ and the Fermi level $E_F$ oscillate because the number of donors in the depletion layer is not infinite. The calculation assumes that DOS between LLs vanishes. Figure 4 shows the calculated electron density in GaAs QW for two spacer values. The calculations were carried assuming parabolic (P) or nonparabolic (NP) energy band in GaAs. The second assumption is more realistic but this point is not essential for our purposes. Two important features should be emphasized. First, for the smaller spacer width $L_S$, the 2D electron density $N$ at $B = 0$ is higher, see also Ref. [27]. Second, for the smaller spacer the oscillations of $N$ have a distinctly higher amplitude. Figure 5 shows corresponding results for the behaviour of Fermi level calculated for the smaller spacer $L_S$ = 8.2 nm. It is seen that, while $E_F$ ($B$) also oscillates, the drops on the higher field sides are not vertical. As explained above, this feature is due to the increase of $N$ at the field values corresponding to $E_F$ between LLs, see Fig. 3(b).

## 3 Quantum transport

**3.1 Quantum magneto-transport** In this section we consider magneto-transport effects from the point of view of the reservoir model. Historically, this subject is of central importance because the very idea of a reservoir was conceived by Baraff and Tsui [2] in relation to the Quantum Hall Effect. Also, there exists huge literature concerned with the explanation of this phenomenon. However, in our present perspective the quantum transport is just another important physical domain in which the existence of a reservoir can be manifested. One should also be aware that, in general, experiments in transport phenomena are not easy to interpret because many physical factors come simultaneously into play. We begin by a simple description of experimental results on QHE and then review briefly other papers related to this subject.

According to the classical Drude model the conductivity components for the degenerate electron gas are

$$\sigma_{xx} = \frac{e^2 N}{m^*}\frac{\tau}{1+\omega_c^2\tau^2}, \quad \sigma_{xy} = -\frac{e^2 N}{m^*}\frac{\omega_c\tau^2}{1+\omega_c^2\tau^2} \tag{7}$$

where $\tau$ is the relaxation time. The above relations give

$$\sigma_{xy} = -\frac{ecN}{B} + \frac{\sigma_{xx}}{\omega_c\tau} \tag{8}$$

and

$$\frac{\sigma_{xx}}{\sigma_{xy}} = -\frac{1}{\omega_c\tau}. \tag{9}$$

In the range of high fields $\omega_c\tau >> 1$ one has for arbitrary degeneracy of 2DEG

$$\sigma_{xy} \approx -\frac{ecN}{B}. \tag{10}$$

One can introduce the well known resistance tensor with the components

$$\rho_{xx} = \frac{\sigma_{xx}}{\sigma_{xx}^2+\sigma_{xy}^2}, \quad \rho_{xy} = \frac{-\sigma_{xy}}{\sigma_{xx}^2+\sigma_{xy}^2}. \tag{11}$$

For strong degeneracy one has at all fields

$$\rho_{xy} \approx \frac{-1}{\sigma_{xy}} = \frac{-B}{ecN}, \tag{12}$$

while at high fields one has for arbitrary degeneracy

$$N = \frac{\rho_{xy} B}{ec\left(\rho_{xx}^2+\rho_{xy}^2\right)}. \tag{13}$$

The above relations can be used to determine the electron density from magneto-transport experiments. However, it is clear that, in the standard interpretation in which $N$ = const and the plateaus of the quantum Hall are attributed to the localization regions of DOS, formula (12) *can not* be used in the plateau region because it would not give a constant $\rho_{xy}$ when $B$ increases and $N$ is constant.

The LL broadening parameter $\Gamma$ is related in case of short-range scattering to the zero field relaxation time $\tau$ [28]

$$\Gamma = \left[\frac{2}{\pi}\hbar\omega_c\frac{\hbar}{\tau}\right]^{1/2} = \sqrt{\frac{2}{\pi}\frac{\hbar e}{m^* c}}\left(\frac{cB}{\mu}\right)^{1/2}, \tag{14}$$

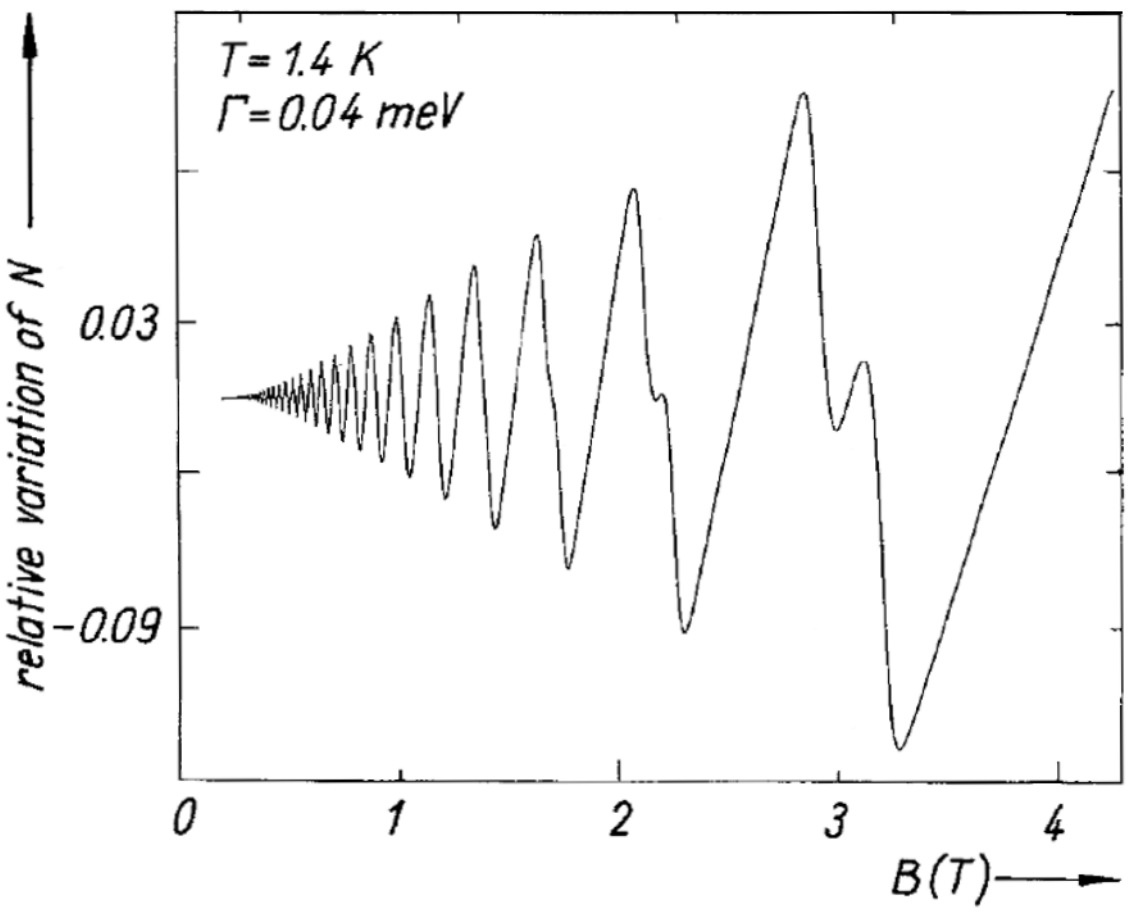

**Figure 6** Relative variation of $N$ versus magnetic field in GaAs QW (sample 2) as calculated using the reservoir model [$N(0) = 3.7 \times 10^{11}$ cm$^{-2}$]. After Ref. [10].

where $\mu$ is the carrier mobility. Finally, if one assumes the Gaussian DOS for LLs, the longitudinal conductivity $\sigma_{xx}$ is, in the high field regime $\omega_c\tau >> 1$ [28]

$$\sigma_{xx} = \frac{e^2}{\pi^2\hbar}\times \int_{-\infty}^{\infty} \frac{-\partial f(E)}{\partial E} \sum_{n,s}\left(n+\frac{1}{2}\right)\exp\left[-\left(\frac{E-\lambda_{ns}}{\Gamma}\right)^2\right]dE. \quad (15)$$

However, the above formalism does not include the localisation of electron states. Thus, in order to describe the plateaus of $\rho_{xy}$ and zeros of $\rho_{xx}$ we use the reservoir hypothesis following the work of Raymond and Sibari [10]. This treatment uses the triangular well approximation. The electron density is calculated according to Eq. (2) with the fixed Fermi energy $E_F$.

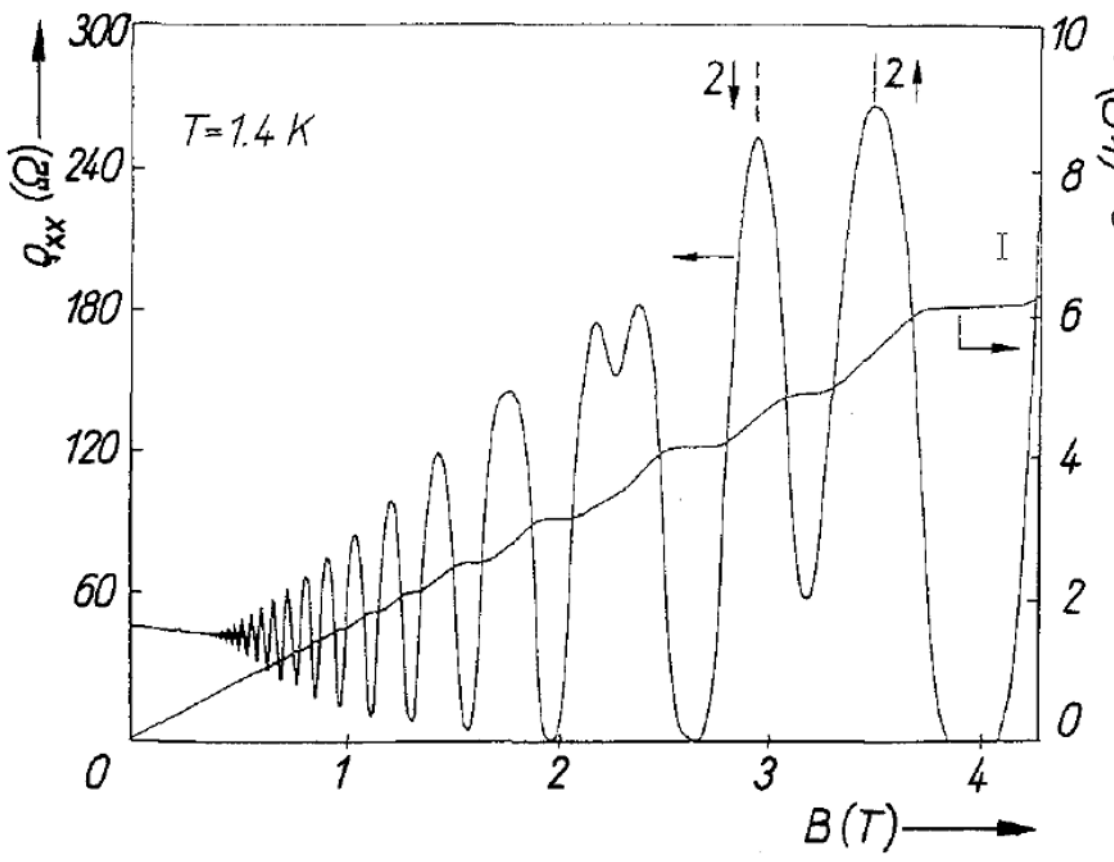

**Figure 7** Experimental dependences of $\rho_{xx}$ and $\rho_{xy}$ on the magnetic field for sample 2 of GaAs/GaAlAs heterostructure. After Ref. [10].

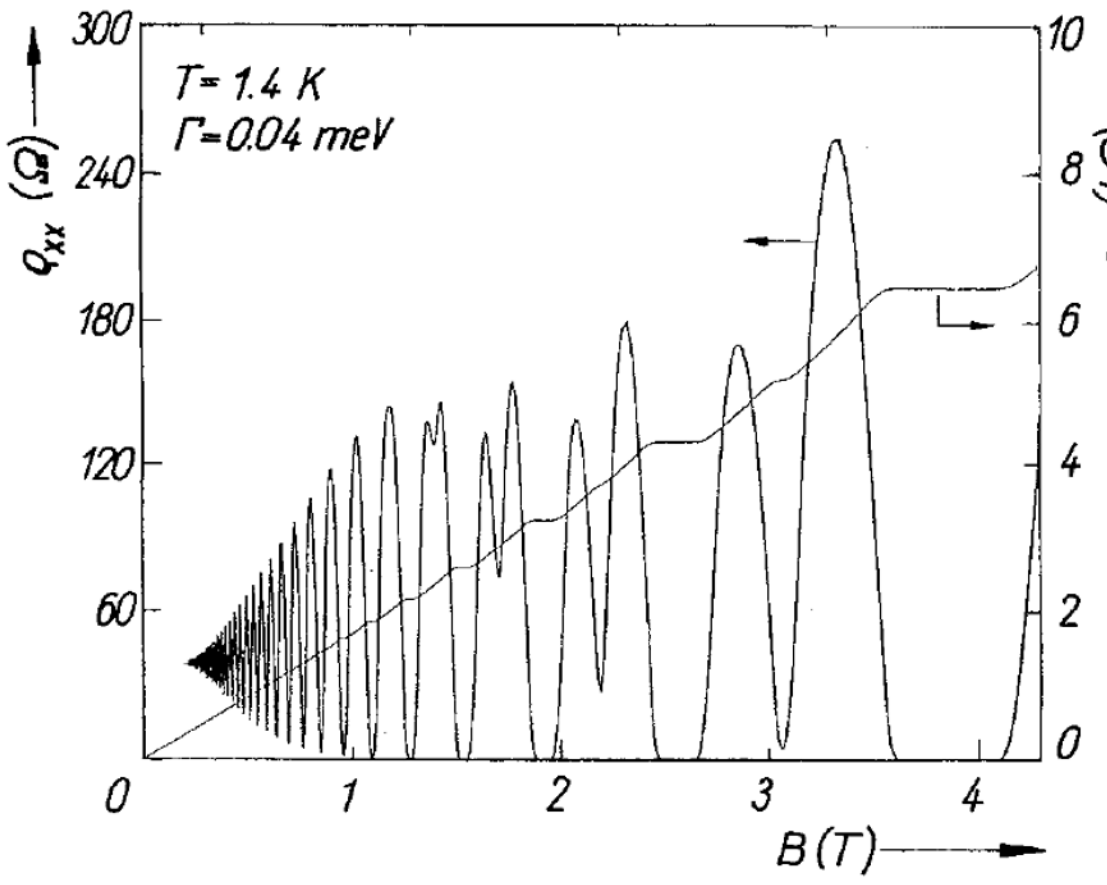

**Figure 8** Theoretical dependence of $\rho_{xx}$ and $\rho_{xy}$ on the magnetic field for sample 2 shown in Fig. 7, calculated assuming a fixed value of the Fermi energy. After Ref. [10].

The value of spin $g$ factor appearing in Eq. (2) should take into account the exchange enhancement [29]. There is $g^* = g_0^* + \Delta g$, where

$$\Delta g = \frac{hc}{eB}\frac{2(N\uparrow - N\downarrow)}{\sqrt{2n+1}}\frac{m_0}{m^*}\left[1+\frac{4\pi\varepsilon L\hbar\omega_c}{e^2}\right]^{-1}. \quad (16)$$

The symbols $N\uparrow(\downarrow)$ represent the densities of mobile electrons having $\uparrow(\downarrow)$ spins and $\varepsilon$ is the dielectric permittivity of GaAs.

We showed in the previous section that, if $E_F$ remains constant as $B$ varies, the density $N$ oscillates as a function of magnetic field. This corresponds to the transfer of electrons into and out of quantum well. The conductivity $\sigma_{xx}$ is calculated according to Eq. (15), while $\sigma_{xy}$ is determined from Eq. (10) once the oscillating $N$ is calculated. Finally, $\rho_{xx}$ and $\rho_{xy}$ are deduced from Eq. (11). The experimental data were obtained on GaAs/GaAlAs heterojunctions grown by MBE and MOCVD techniques. The samples had different values of the density $N$ and carrier mobility $\mu$. These parameters were also changed by applying a hydrostatic pressure and by illuminating the samples with infrared light emitting diodes. It was found that the best description of the data was obtained for the broadening parameter $\Gamma$ determined by the zero-field mobility $\Gamma \approx \hbar/\tau \approx \hbar e/m^*\mu$.

Figure 6 shows the relative variation of 2D density $N$ in GaAs QW calculated assuming the constant Fermi energy. The exchange enhancement of the spin $g$ value is clearly seen for $B \approx 3$ T. Figures 7 and 8 illustrate experimental and theoretical values of $\rho_{xx}$ and $\rho_{xy}$ for one of the investigated GaAs/GaAlAs heterostructures. The following parameters were used in the calculations: $m^*/m_0 = 0.07$, $\varepsilon = 12.91$, $N_A - N_D = 2\times10^{14}$ cm$^{-3}$ and the band bending potential $\varphi_d = 1.425$ eV. It can be seen that the agreement between experiment and theory is remarkably good.

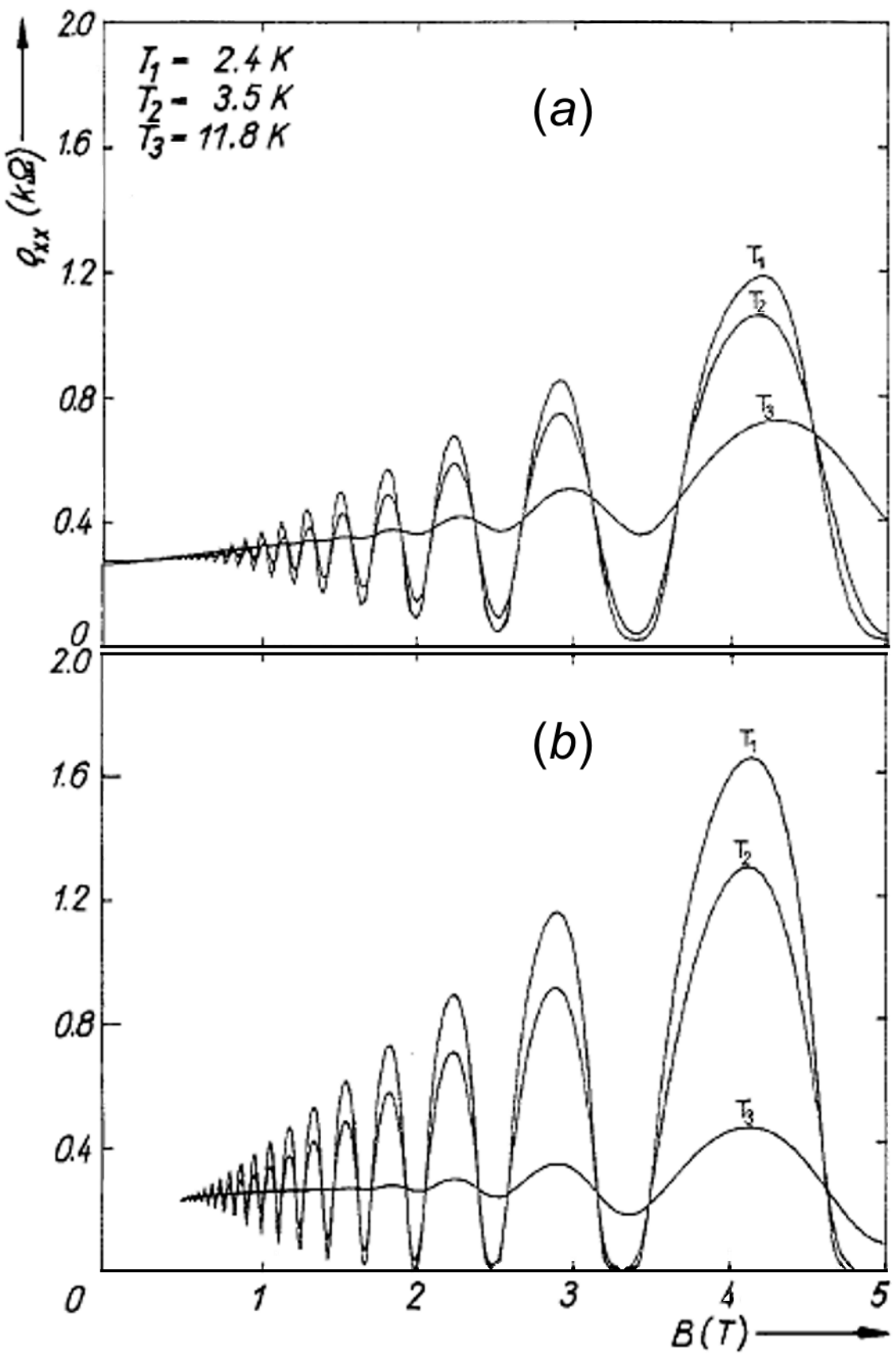

**Figure 9** a) Experimental values of $\rho_{xx}$ versus $B$ for low mobility sample 5 at three temperatures. b) Corresponding theoretical values for sample 5, calculated assuming a fixed value of the Fermi energy. After Ref. [10].

A few remarks are in order. The agreement is achieved without adjustable parameters with very narrow Landau levels. The small broadening parameter $\Gamma$ = 0.04 meV is determined from the measured mobility. The plateaus of $\rho_{xy}$ result from the increase of $N$ (see Fig. 6) by the mechanism explained in the previous section. The corresponding zeros of $\rho_{xx}$ result from the fact that the Fermi level $E_F$ is at these $B$ values between the levels, where the density of states has been assumed to vanish. One should bear in mind that the correct values of the quantum Hall plateaus are assured automatically by the degeneracy of Landau levels, while the linear increase of $N$ with a magnetic field (seen very well in Fig. 6) is assured by the proportionality of the total DOS to $B$, see Eq. (1). Thus the measure of agreement between the experiment and theory is the coincidence as a function of $B$ and it is truly good. On the critical side, the theoretical spin splitting due to the exchange enhancement of $g$ value is larger than that observed experimentally. This could be due either to the theoretical overestimation of the enhancement of $g$ (which is not of our concern here) or to too small value of the broadening parameter $\Gamma$.

In Fig. 9 we quote experimental and theoretical values of $\rho_{xx}$ for a low mobility sample at three temperatures. Again, the agreement is remarkably good. One could have a still better agreement taking a somewhat smaller value of $\Gamma$. The value employed in the calculations was determined from the mobility at $B$ = 0. In addition, quite a good description of the Quantum Hall Effect in samples subjected to various hydrostatic pressures up to 11.3 Pa was obtained using the same approach, see Ref. [10].

However, the description of magneto-quantum transport at temperatures below 1 K with the same reservoir approach is not so successful. In order to reach a satisfactory agreement between experiment and theory one needs additional assumptions, which are not well justified. Thus, in order to describe correctly the low temperature data like, for example, those of Ebert et al. [30], one manifestly needs to evoke the electron localisation. This supports our previous statement that the transport phenomena are more complicated to interpret.

Now we briefly mention other work concerned with the reservoir approach to quantum magneto-transport. The pioneering work of Baraff and Tsui [2] contained all the essential results of the reservoir approach. It used the self-consistent procedure for describing the electron transfer between the depletion layer in the GaAlAs barrier and GaAs quantum well showing that this approach gave the correctly quantized plateaus of Hall resistance. The obtained plateaus were somewhat too narrow compared to experimental data. The paper of Bok and Combescot [3], using basically the same procedure, calculated in addition capacitance oscillations in the junction. The authors made an observation that the capacitance is sensitive to the total DOS (both localized and delocalized), so that a comparison with the transport data may be used to determine the amount of localized states, cf. our Fig. 1. This idea was used later in relation to the behaviour of Fermi level and magnetization, see below. Konstantinov et al [8] considered theoretically a metal-oxide-semiconductor structure with a reservoir of surface states at the insulator- semiconductor interface and obtained for $T$ = 0 a sequence of quantum Hall plateaus. Toyoda et al [31] in a non-selfconsistent consideration attempted to explain widths of the quantum Hall plateaus observed by Störmer et al [32] by putting upper and lower limits on the electron transfer from the reservoir. Raymond and Karrai [6] obtained a good description of their QHE data by assuming that the Fermi level was completely fixed by a reservoir at the GaAs/GaAlAs interface, see our Figs. 7 and 8. An almost equally good description was also obtained for experiments under hydrostatic pressure.

Ingraham and Wilkes [9] considered a fixed Fermi energy and showed that it leads to a correct description of experimental quantum magneto-transport data of various authors at low temperatures when 2DEG is degenerate. The authors concluded that the reservoir must have electron states at all energies if it is to act as a source or sink of electrons both in the rises and in the plateaus of QHE. Xu

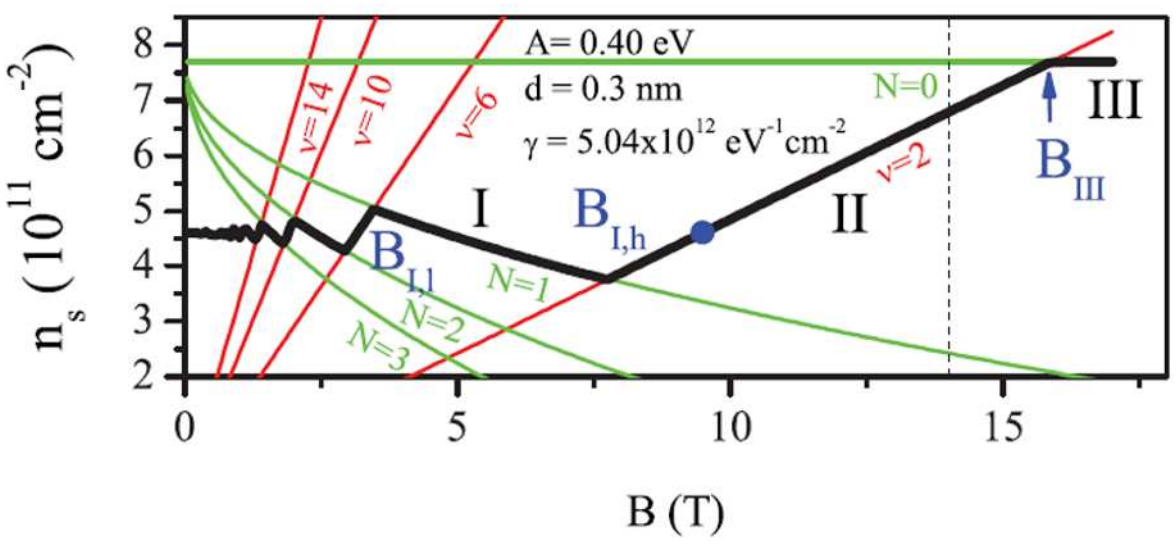

**Figure 10** Carrier density in monolayer graphene grown on SiC versus magnetic field, as described by the charge transfer model (black line). For details see Ref. [19].

[24] carried a self-consistent calculation for a GaAs/GaAlAs selectively doped heterostructure at $T = 0$ in the spirit proposed by Baraff and Tsui with some refinements. The magnetic oscillations of the depletion length in the GaAlAs barrier were explicitly displayed. Sabin del Valle and de Dios-Leyva [25] performed a similar calculation for $T > 0$ and two spacer values. Their results for the electron transfer are shown in our Figs. 4 and 5.

A striking example of a very broad QHE plateau related to the charge carrier transfer from a reservoir was recently observed in monolayer graphene grown on SiC, see Janssen et al. [19]. In Fig. 10 we reproduce the calculated change of the 2D density $N(B)$ obtained for the investigated sample. It is seen that, beginning from $B \approx 7.5$ T the density increases linearly with the field up to $B \approx 15$ T. The electrons are provided by surface-donor states in SiC (Si interface). Figure 11 shows the measured magneto-transport components $\rho_{xx}$ and $\rho_{xy}$ versus magnetic field. The very large plateau of $\rho_{xy}$ and the corresponding zero value of $\rho_{xx}$ are seen (highest available magnetic field was $B = 14$ T). According to the interpretation given in Ref. [19] the quantum Hall and ShdH plateaus are solely due to the electron transfer. On the other hand, a very large plateau of QHE (its width was more than 20 T) observed on graphene also grown on SiC (but with C interface) is not believed to be stabilized by a reservoir, see Jouault et al. [33].

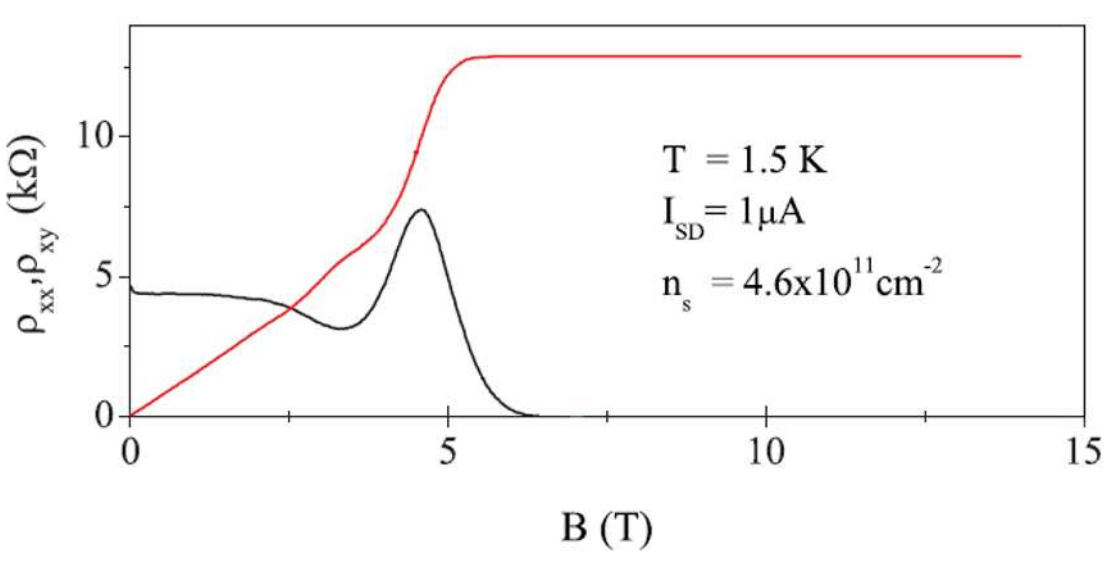

**Figure 11** Experimental $\rho_{xx}$ and $\rho_{xy}$ of monolayer graphene grown on SiC versus magnetic field. A very broad plateau of QHE is seen, corresponding to the linear increase of charge density shown in Fig. 10. After Ref. [19].

Toyoda and Zhang [34] developed the theory of QHE in monolayer graphene in the reservoir model. The authors make a vague statement that “the electron reservoir is the 2DEG itself”. The resulting description agrees well with experimental data of Zhang et al. [35]. Toyoda [36] considered radiation-induced magneto-resistance oscillations in GaAs/GaAlAs heterostructures, first observed by Zudov et al. [37] and Mani et al. [38], and demonstrated that they can be well described by the reservoir model. In particular, this description accounts for the fact that the oscillations are independent of the radiation polarization, as observed by Smet et al. [39].

Finally, we mention two nontypical investigations which used the reservoir model in magneto-transport. Von Ortenberg et al. [40] showed theoretically that the resonant donor state introduced by Fe atoms in the conduction band of zero-gap HgFeSe serves as an electron reservoir and can lead to 2D-like behaviour of magneto-transport tensor when the samples are made sufficiently thin. Kulbachinski et al. [41] investigated experimentally bulk semimetal alloys $Bi_2Te_3$ and $Sb_2Te_3$ demonstrating that an overlap of the conduction and valence bands, which works effectively as an electron reservoir, leads to plateaus of QHE.

**3.2 Oscillations of Fermi level and thermoelectric power** We showed in Section 2 that the Fermi level also oscillates as the magnetic field $B$ changes. Such oscillations can be measured by the so called floating-gate technique. We do not go into explanations of this method but show the results. The behaviour of $E_F(B)$ is of importance since, as follows from the figures in Section 2, it can shed light on the subject of our interest. Namely, as follows from Fig. 2(a), in the regime of constant density $N$ the drops of $E_F$ on the higher-field sides are vertical if there is no background DOS between LLs. On the other hand, Figs. 4 and 5 indicate that in the regime of oscillating $N$ and $E_F$ the drops of the Fermi level on high-field sides are not vertical.

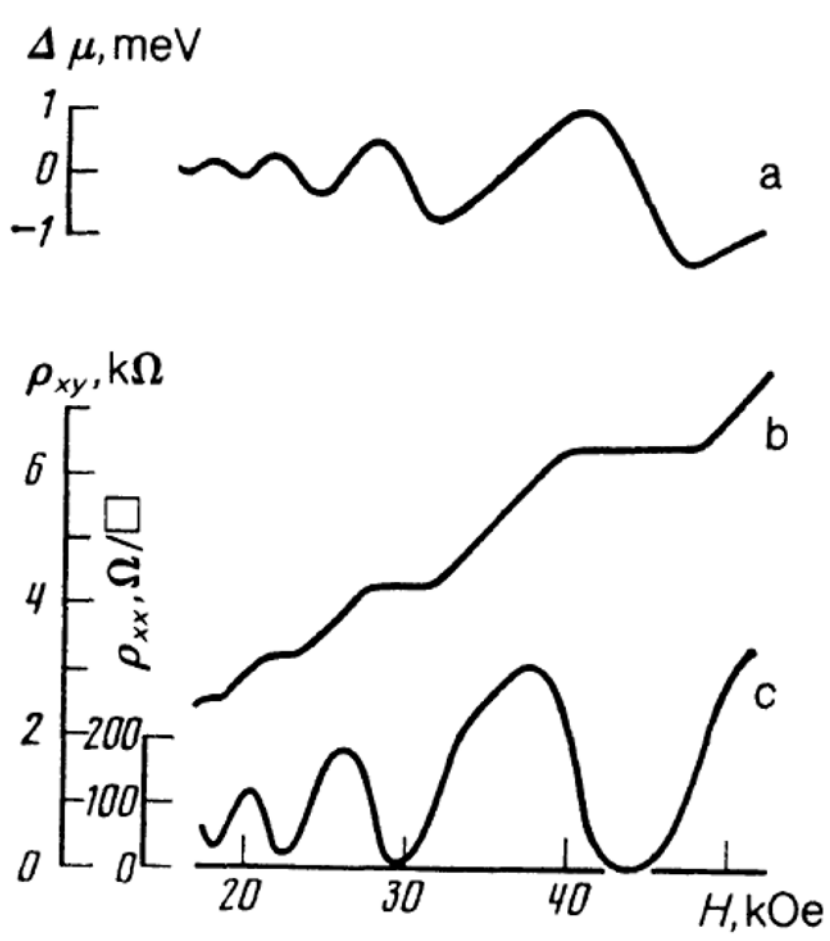

**Figure 12** The chemical potential (a), Hall effect (b), and magneto-resistance (c) versus magnetic field, measured on a GaAs/GaAlAs heterojunction. After Ref. [11].

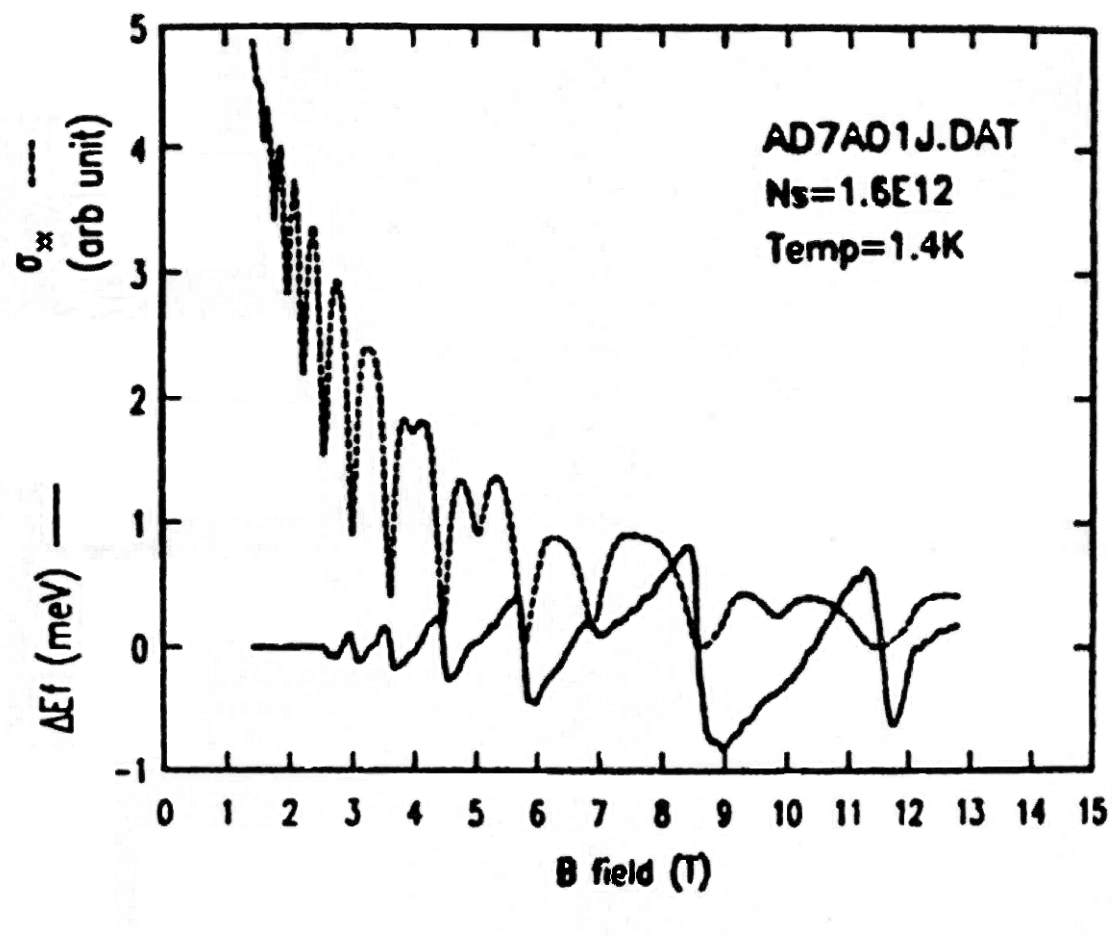

**Figure 13** The chemical potential and conductivity $\sigma_{xx}$ versus magnetic field, measured on a Si inversion layer. After Ref. [42].

Figure 12 shows in three parts the experimental behaviour of Fermi energy (a), Quantum Hall Effect (b), and ShdH effect (c) in GaAs/GaAlAs heterostructure measured by Nizhankovskii et al. [11]. It is seen that the drops of $E_F$ on the high-field sides of oscillations are far from vertical. This can correspond to either background DOS between LLs or to the 2D electron transfer in the reservoir approach. The authors attributed their results to the electron transfer and showed that the latter can explain the QHE data. In Fig. 13 we quote results of Fang et al. [42] on $E_F$ oscillations and the ShdH effect observed on a Si inversion layer. It is seen that the drops of $E_F$ on high-field sides are also not vertical. A direct comparison of $E_F$ with transport effects can furnish another useful information. The behaviour of Fermi energy is not sensitive to the mobility edges between localized and nonlocalized electron states. On the other hand, the ShdH effect is nonzero only when $E_F$ is in the delocalized region of DOS, see Fig. 1. This means that the zeros of ShdH effect should be somewhat larger on the $B$ scale than the high-field drops of $E_F$. The data shown in Fig. 13 do not show this feature suggesting that the background DOS between LLs and the localization regions on the shoulders of LLs is practically vanishing in the investigated sample.

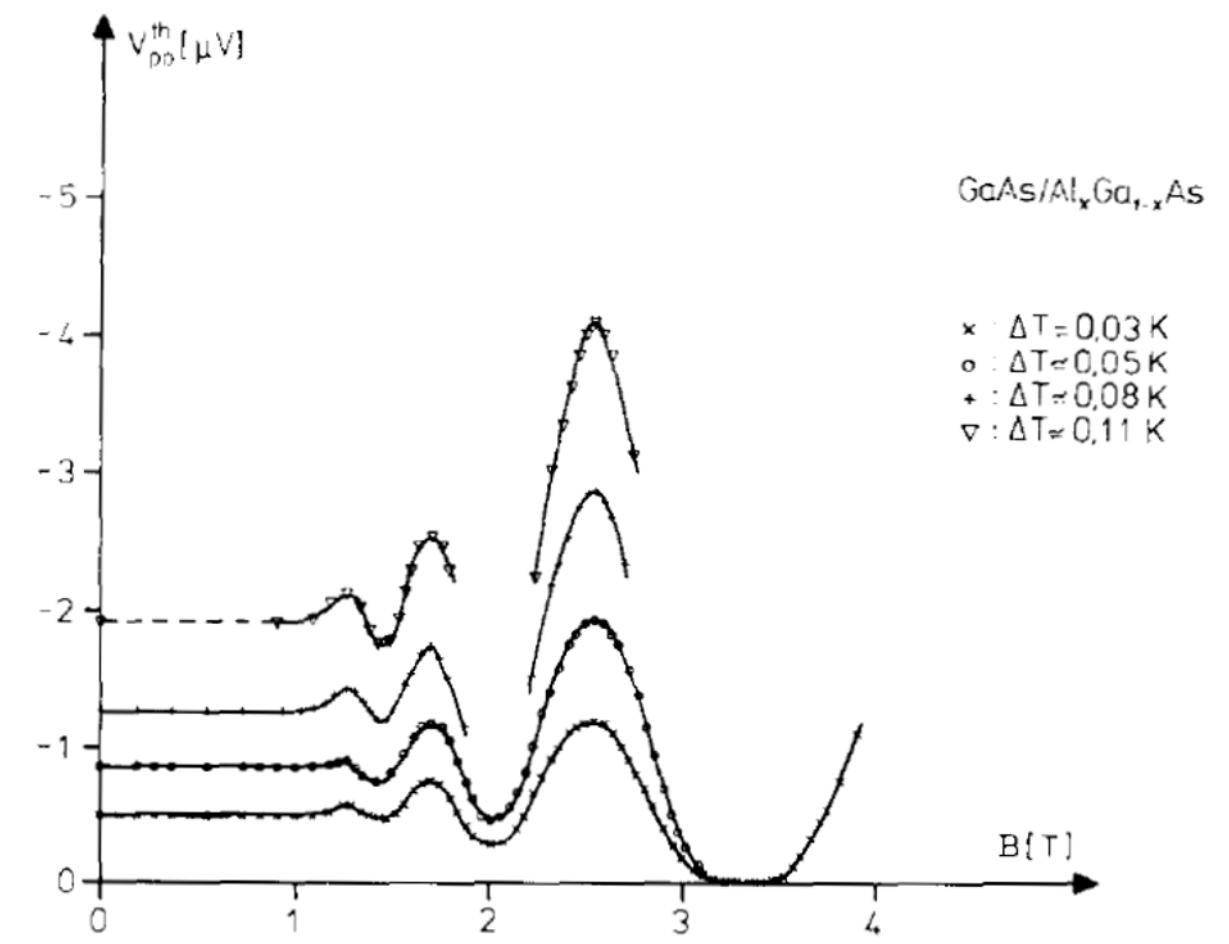

**Figure 14** Thermal voltage versus magnetic field at different temperature gradients $\Delta T$, measured on a GaAs/GaAlAs heterostructure at $T = 4.2$ K. After Ref. [43].

Figures 2(c) and 3(e) in Section 2 illustrate the behaviour of thermoelectric power in a magnetic field $\alpha(B)$ in the regimes of constant $N$ and constant $E_F$, respectively. In Fig. 14 we quote experimental results of Obloh et al. [43] on $\alpha(B)$ in GaAs/GaAlAs heterostructure. It is seen that the "plateau" of vanishing thermoelectric power for fields above 3 T strongly resembles the result shown in Fig. 3(e) corresponding to the electron transfer. Still, according to the standard interpretation this plateau can be alternatively explained by the localized states in the LL density of states.

**4 Cyclotron resonance** It is clearly of interest to investigate 2DEG in the quantum Hall regime, and in particular its density $N$, by means alternative to the magneto-transport. In particular, the 2D electron gases in heterostructures can be studied by the cyclotron resonance (CR). An important property of the CR phenomenon is that it involves electrons located in both localized and extended regions of the Landau levels. Manasreh et al. [12] were the first to demonstrate that CR can be used to determine the density of 2DEG in the conditions of QHE and they showed by measuring the total integrated absorption that $N$ oscillates as the magnetic field is varied (their Fig. 6). These studies were extended by Raymond et al. [13] in two ways. First, it was attempted to correlate the CR data with the transport data. Second, the authors explained the CR data by means of the reservoir model. Below we briefly review this investigation.

One can describe the light transmission in the presence of a magnetic field (the Faraday configuration) using the Drude-type model. It is assumed that the Kohn theorem is valid, so that $m^*$ is not sensitive to many-body effects. The resulting conductivities for left and right circularly polarized radiation are

$$\sigma_{\pm}(\omega) = \frac{Ne^2\tau}{m^*}\frac{1}{1+i(\omega\pm\omega_c)\tau}, \qquad (17)$$

where $\omega$ is the radiation frequency and other quantities were defined above. One can express the transmission of linearly polarized light by $\sigma_{\pm}$. In the QHE regime one can deal with one or two CR-type transitions (neglecting spin), depending on the position of the Fermi energy, cf. Fig. 15. Suppose first that $n$ Landau levels are fully occupied by the electrons and the higher ones are empty. The quantum mechanical probability $C$ of the $n$ to $n + 1$ transition is

$$C = \frac{2eB}{hc}(n+1)\delta(\omega_c - \omega), \tag{18}$$

where $\delta$ is the Dirac delta function. Since $2eB/hc$ is the degeneracy of an LL in 2D system (the spin splitting is neglected), one can replace $(2eB/hc)(n + 1)$ by $N$ in the classical description. Suppose now that the Fermi level is within the $n + 1$ level and the filling factor of this level is $\nu = \alpha$ (cf. Fig. 15). By taking into account the filling factors, the complete probability of the two CR transitions is

$$\begin{aligned} &\frac{2eB}{hc}[(n+1)(1-\alpha)+(n+2)\alpha] = \\ &= \frac{2eB}{hc}(n+1+\alpha) = N. \end{aligned} \tag{19}$$

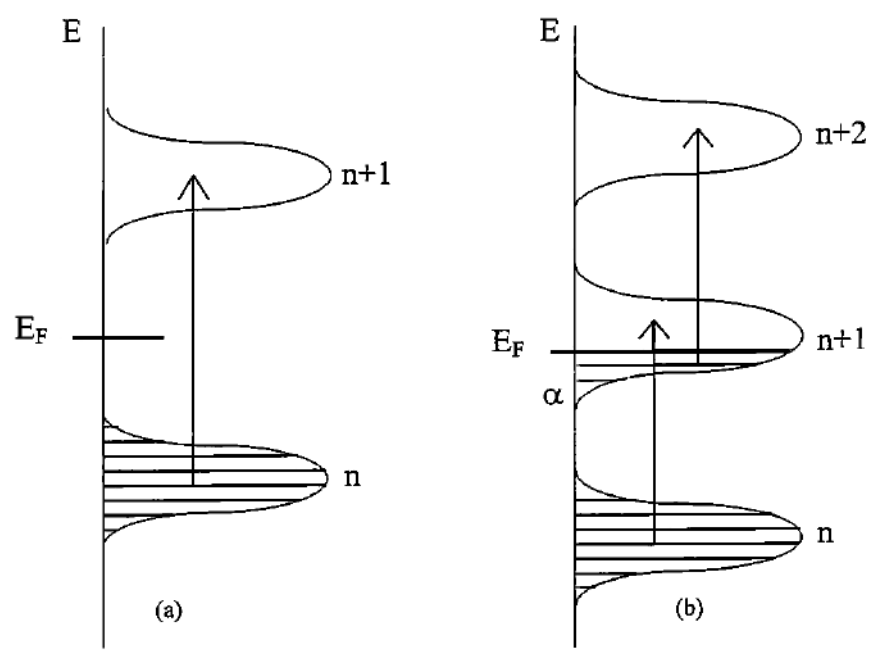

**Figure 15** Cyclotron resonance transitions in a 2D system: (a) for the Fermi energy located between Landau levels and (b) for $E_F$ located within a Landau level (schematically). The filling factor of the $n + 1$ level is $\alpha$. After Ref. [13].

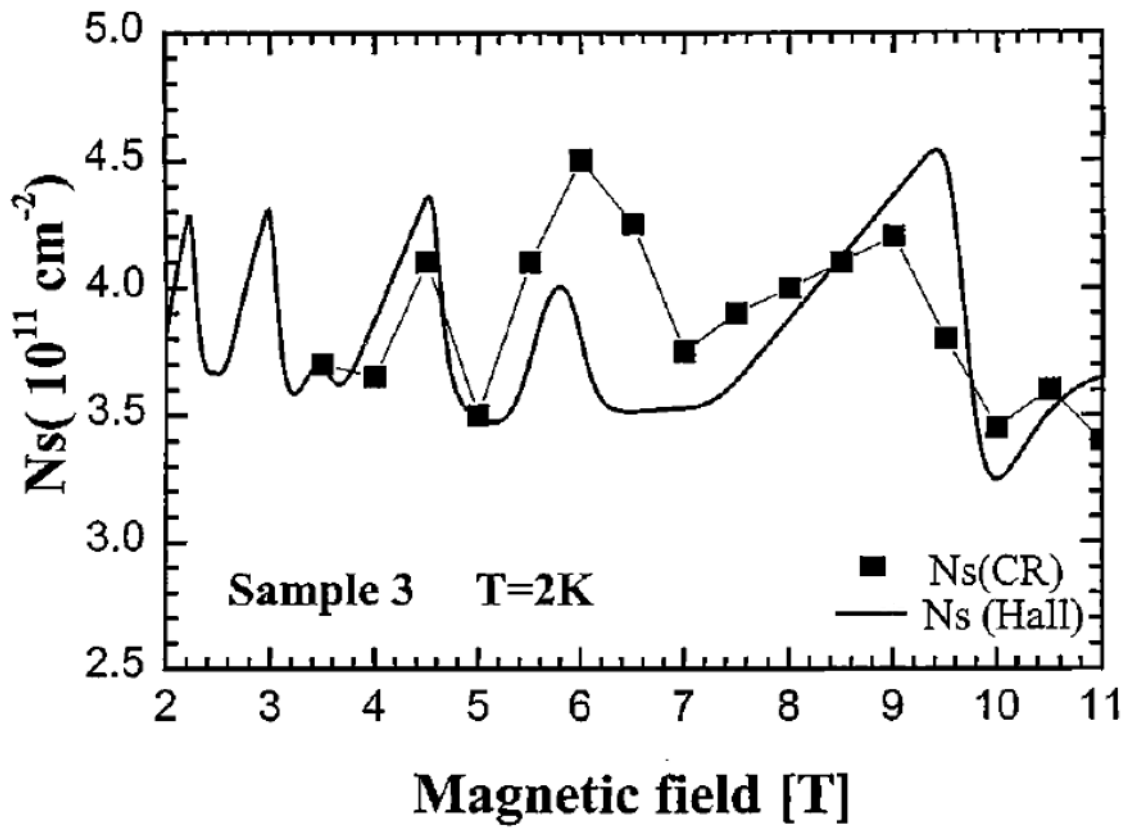

**Figure 17** The same as in Fig. 16 but for sample 3. After Ref. [13].

Thus, also in case of the partial occupation of the level, when two CR transitions are possible, the complete probability may be expressed by $N$ in the classical description. This reasoning is valid for a 2D system with its “universal” degeneracy of Landau levels, but not for a 3D system.

The CR measurements and the quantum transport measurements were performed on the same $GaAs/Ga_{0.67}Al_{0.33}As$ heterostuctures in the QHE regime. The best fit to the CR data was used to determine two parameters: $\tau$ and $N$, see Eq. (17). The relation $\rho_{xy} = B/Nec$ was used to determine the electron density $N$ from the Hall effect at all magnetic fields. As mentioned above, in this interpretation in order to have a plateau of $\rho_{xy}$ the density $N$ must increase linearly with $B$.

Figures 16, 17, and 18 summarize the main findings. It is seen in Figs. 16 and 17 that both the cyclotron resonance and $\rho_{xy}$ measured in the Hall effect indicate oscillatory character of the electron density $N$ with varying magnetic

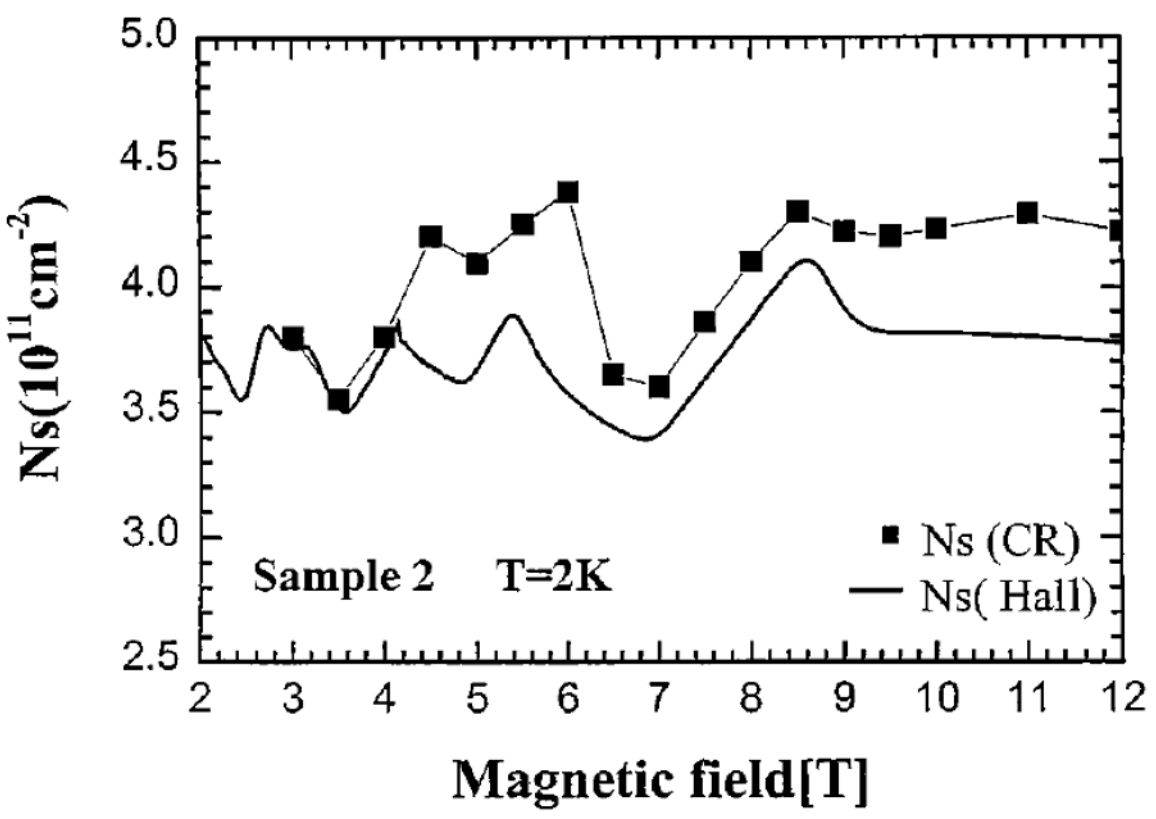

**Figure 16** The electron density for GaAs/GaAlAs (sample 2) versus magnetic field, as determined by a fit to the cyclotron resonance data and from the Hall resistance with the use of the formula $\rho_{xy} = B/Nec$. After Ref. [13].

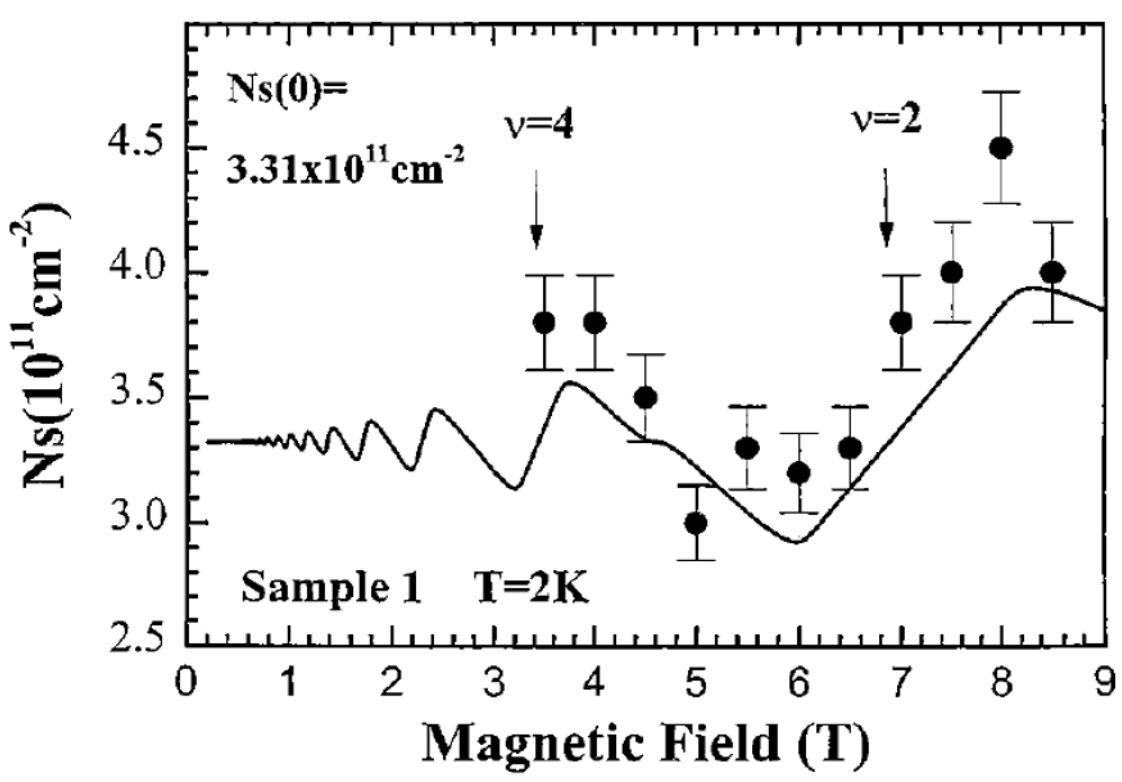

**Figure 18** The electron density for GaAs/GaAlAs (Sample 1) versus magnetic field, as calculated using the reservoir model (solid line). The assumed zero-field value is measured by the Hall effect. Arrows indicate the filling factors. Full circles are the experimental CR values. After Ref. [13].

field for two different samples. In Fig. 18 the electron density determined from the CR data is compared with calculations assuming that an outside reservoir pins the Fermi level. It is seen that the reservoir hypothesis works quite well. Two important conclusions of the above analysis are: 1) The electron density $N$ in GaAs QW oscillates as a function of a magnetic field, 2) The "classical" formula $\rho_{xy} = B/Nec$ seems to work at all fields, also in the quantum Hall regime.

**5 Magneto-photo-luminescence** From the early days of optical experiments with semiconductor heterostructures it was observed that the energies of interband magneto-photo-luminescent (MPL) transitions exhibit striking nonlinear behaviour as functions of an external magnetic field. Such nonlinearities are characteristic of 2D systems and are not seen on bulk materials. The investigated systems can contain only one or more populated electric subbands. In the second case, it was shown that the nonlinearities were related to an electron transfer between the subbands. The situation is different if only one subband is populated. An important example of such a situation is a rather narrow and not strongly doped GaAs/$Ga_{0.67}Al_{0.33}$/As quantum well. In this case one cannot explain the nonlinearities by the above mechanism. A typical example of nonlinearities in MPL data is presented in Fig. 19. Two theoretical calculations for this situation were proposed and reached similar conclusions [44, 45]. As the consecutive Landau levels cross the Fermi energy in an increasing magnetic field, the oscillatory density of states gives rise to oscillations of screening. The oscillations of screening result in the oscillatory renormalization of the energy gap which is reflected in the interband energies. For symmetric QWs, the MPL energies should show positive cusps at even filling factors. Tsuchiya et al. [46] extended this work to asymmetric QWs which allow one to separate in the real space electrons and holes. In the theoretical work based on the oscillations of screening the comparison of the theory with experimental data was not convincing. Experimentally, the observed MPL peaks did not occur at even filling factors, they never appeared in the form of cusps and there was no evidence in the literature for the phase reversal of peaks for different well widths, as predicted in Ref. [46].

For these reasons the problem of interband MPL nonlinearities was reconsidered by Zawadzki et al. [14] with the use of the reservoir model (see also Kamal-Saadi et al. [47]). Below we briefly summarize the main points of this analysis. The experiments were performed on asymmetric modulation-doped GaAs/$Ga_{0.67}Al_{0.33}$As QWs of different widths. In such structures electrons in the conduction subband and holes in the valence subband are spatially separated. The considered MPL free-electron transition occurs between $0^+$ ($c$) and $1\beta$ ($hh$) Landau levels and it is marked $F$ in Fig. 19. The main and only assumption is that, because of an external reservoir, the Fermi energy remains constant, so that, as the magnetic field changes, the electrons are transferred between the GaAs conduction QW and the reservoir. The $N(B)$ oscillations cause periodic modification of the self-consistent electric potential. This, in turn, changes the conduction and valence subband energies and results in the nonlinearities of the interband MPL energies. The important feature is the exchange enhancement of the spin $g$ factor when the Fermi energy occurs between two spin levels. The enhanced $g$ value is included self-consistently in a sense that it both provokes and is affected by the electron density oscillations. The $g$ enhancement occurs in the vicinity of the Fermi energy, but this mechanism affects the electron transfer which, via the change of the confining potential, is reflected in the behaviour of all levels.

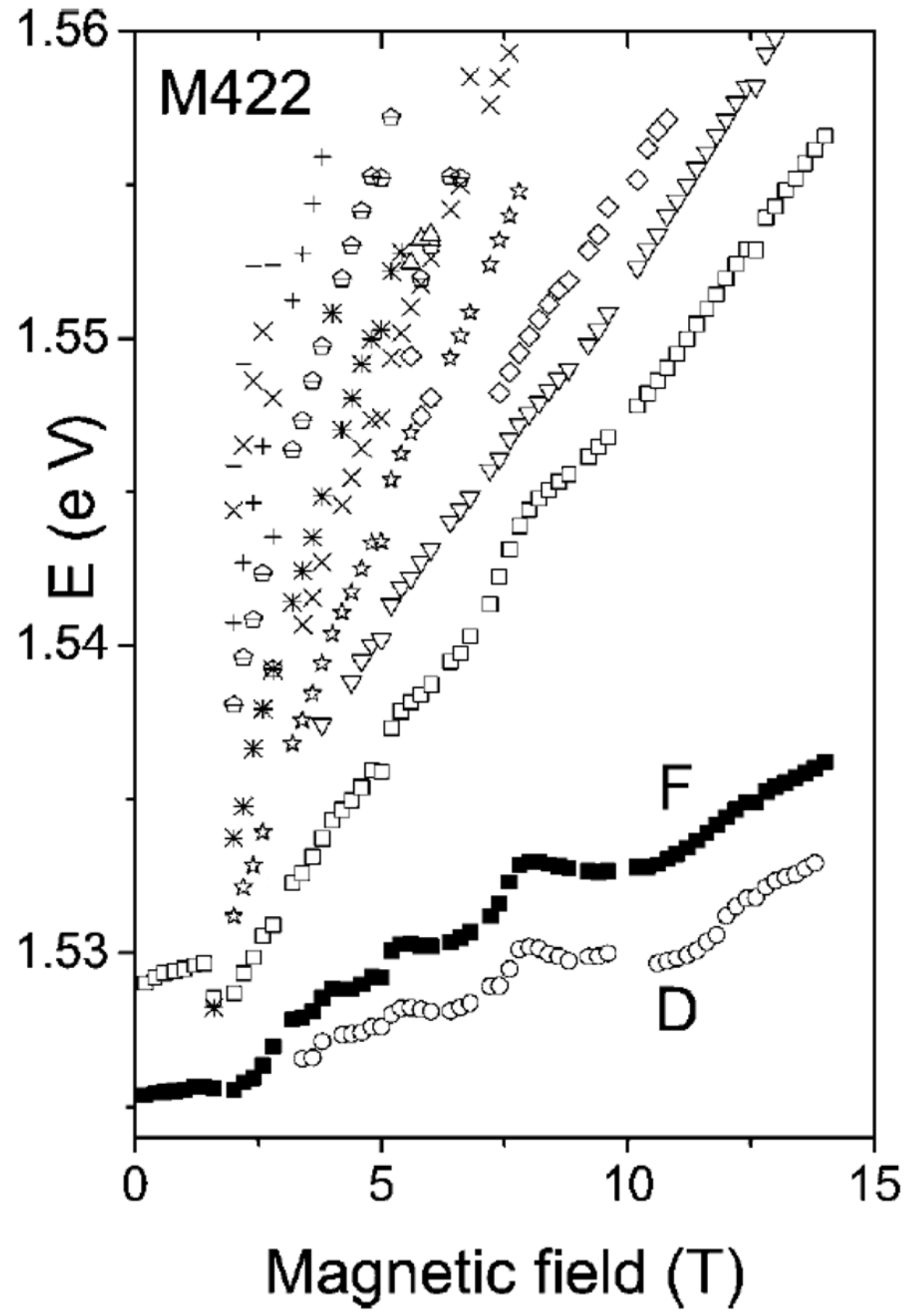

**Figure 19** Fan chart of experimental magneto-photo-luminescence (MPL) energies measured on a GaAs/GaAlAs heterostructure versus magnetic field. The lowest transition is due to donors, the transition $F$ is considered in the theory. After Ref. [14].

The oscillations of $N(B)$ calculated in a selfconsistent approach, the exchange enhancement of the spin $\Delta g$ value near the Fermi level and the energies of the conduction and valence Landau levels in question are shown in Fig. 20(b). The interband MPL energies are given by differences of the conduction and valence energies of LLs. It was found that the comparison between experimental and theoretical

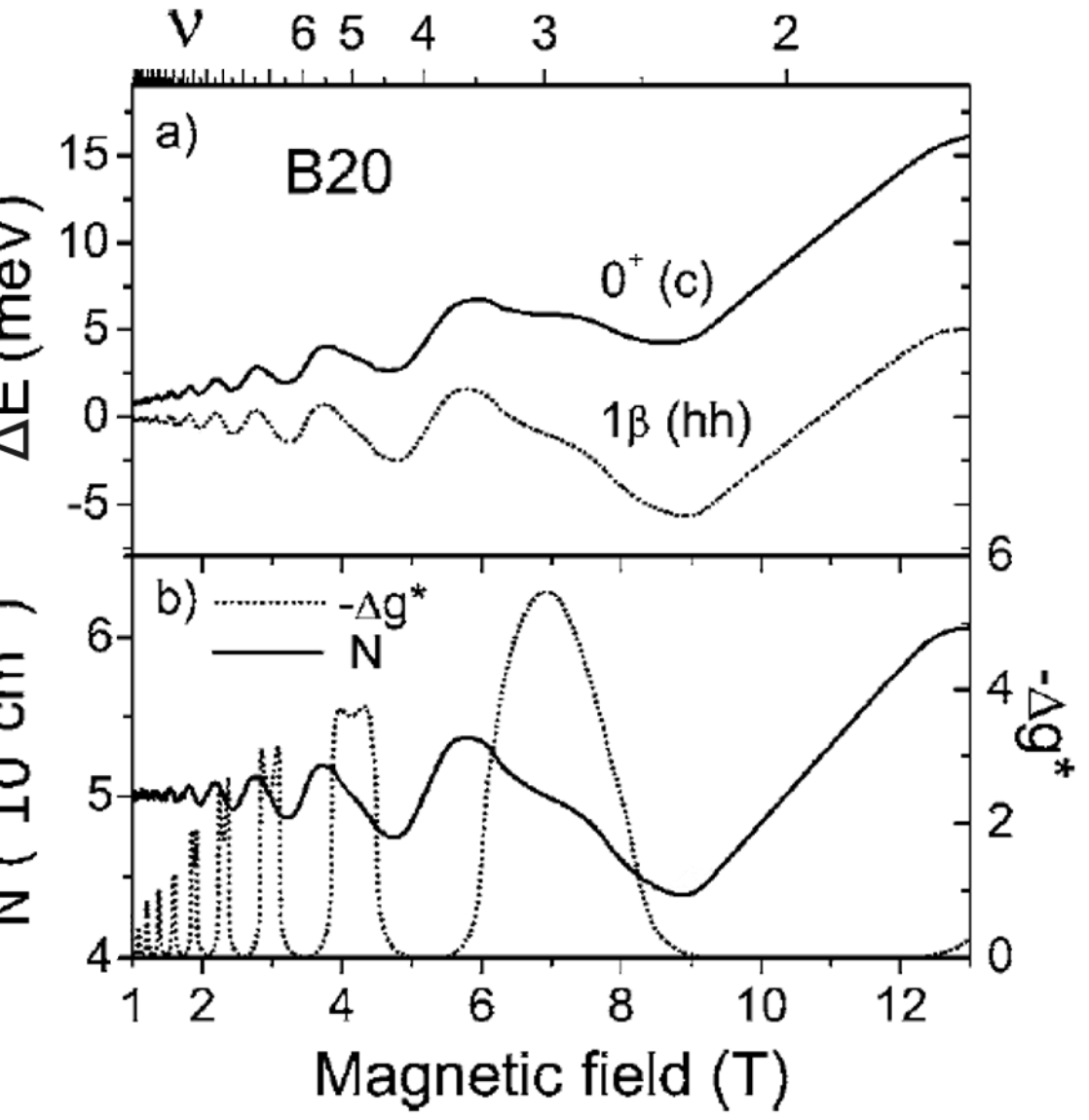

**Figure 20** (a) Energy shifts of the conduction level $0^+$ (*c*) and the heavy-hole level $1\beta$ (*hh*) versus magnetic field for GaAs/GaAlAs sample B20. (b) Calculated electron density $N$ (in $10^{11}$ $cm^{-2}$ units) and exchange enhancement of the spin $g$ value versus magnetic field for the same sample. The corresponding filling factors $\nu$ are indicated on the upper abscissa. After Ref. [14].

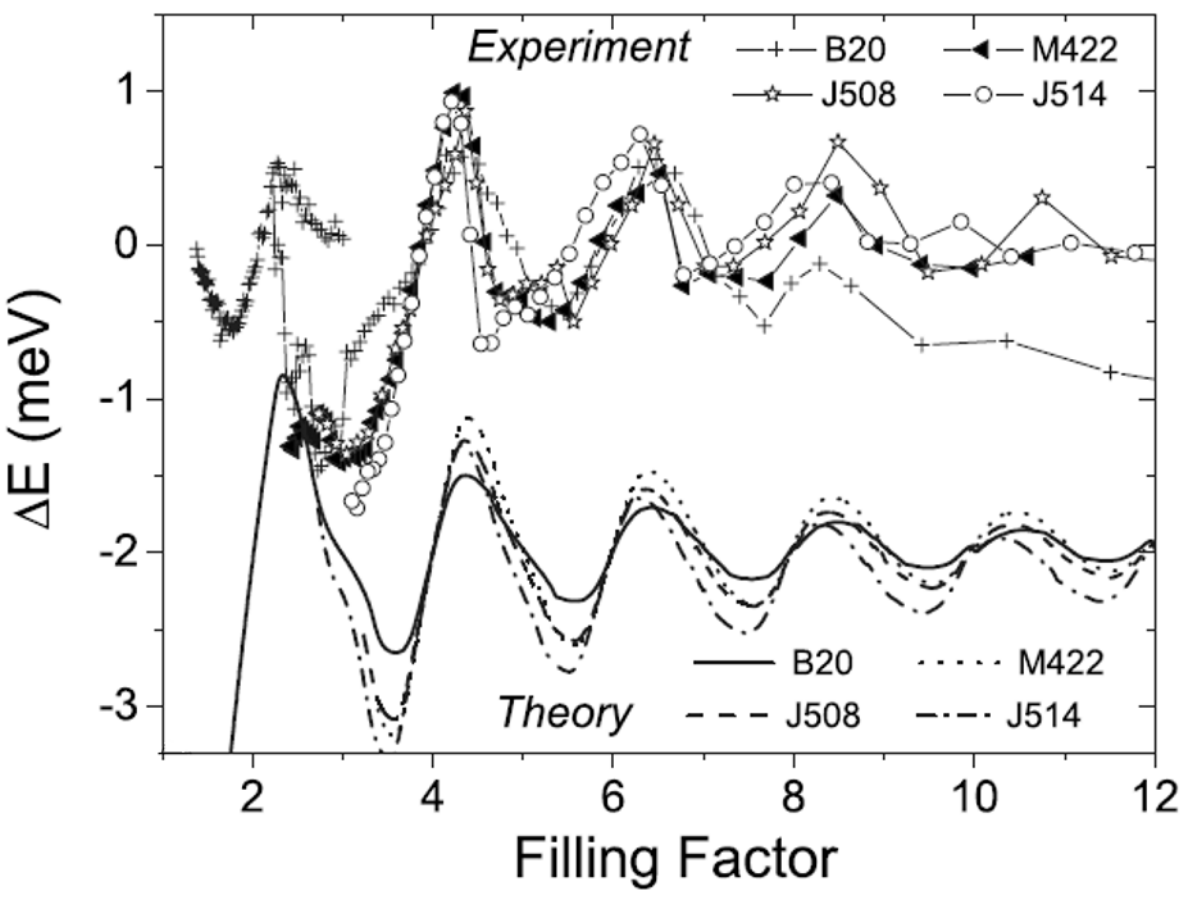

**Figure 21** Oscillatory experimental and theoretical parts of the MPL energies for four GaAs/GaAlAs samples versus filling factor $\nu$. The calculated curves have been shifted down for clarity. After Ref. [14].

energies was more conclusive when the linear $B$ dependences of the oscillating energies were subtracted. The experimental and theoretical findings for four samples having different electron densities are summarized in Fig. 21. In this figure the experiment and theory are plotted as functions of the filling factor $\nu$ assuming that $N$ oscillates. It is seen that the description of the data is very good, both the phases and the amplitudes are well reproduced. Also the details of the description coincide, as discussed in Ref. [14]. It was concluded that the nonlinearities in the PLM energies were caused by the electron density oscillations resulting from the presence of a reservoir. Since, as mentioned above, numerous experiments show the PML energy nonlinearities, the above analysis indicates that many GaAs/GaAlAs structures should be treated as open systems in which the charge transfer between a quantum well and a reservoir is at work.

This conclusion is supported by other investigations using interband optics. In the work of Kukushkin et al. [48] the electron density $N$ in GaAs/GaAlAs hetrostructures was reduced by continuous photo-excitation by laser light. It was found (their Fig. 2) that for different reduced densities $N$, related to different illumination powers, the position of the Fermi energy remained unchanged indicating that its position was stabilized by an external factor. Hayne et al. [49] measured optically induced density depletion of the 2D system at the interface of GaAs/GaAlAs heterojunction. It was shown (their Fig. 3) that, as the laser power varied from 0.1 mW to 20 mW, the position of the Fermi energy remained unchanged. Plentz et al. [50] spatially separated electrons and holes in an asymmetric single-side-doped GaAs/GaAlAs QW structure. The separation was achieved either by an external electric field or by varying the electron density. It was found (their Fig. 4) that for increasing front-gate voltages the electron density $N$ increased while the Fermi energy remained constant. Kerridge et al. [51] studied a system of two QWs created by $\delta$ doping of GaAs/GaAlAs heterostructures: one in the GaAlAs barrier and the other at the GaAs/GaAlAs interface. In this situation the first QW served as an electron reservoir to the second one. It was demonstrated by means of photoluminescence (their Fig. 2) that the electron redistribution between the reservoir and the GaAs QW was strongly affected by changes of an external magnetic field.

**6 Magnetization** In Section 2 we indicated what one can expect oscillatory magnetization of 2DEG in the two limiting regimes of constant electron density and constant Fermi level. The basic feature is that in the regime of a constant electron density the high-field slopes of oscillations are vertical while in the regime of a constant Fermi level the high-field slopes are not vertical. It was also indicated that, in the intermediate regime described by a self-consistent calculation of a modulation-doped heterojunction, the Fermi level also does not drop vertically on the high–field sides. The problem is, however, more complicated since the nonvertical slopes of oscillations can also result from an inhomogeneity of a sample as well as from background DOS between the Landau levels. A distinct feature of the magnetization is that localized and delocalized electron states give similar contributions to the susceptibility, so the dHvA effect is especially well suited for considerations of DOS. As a result, in many investigations an interpretation of the magnetization data was concerned with the form of DOS related to the Landau levels, see the review of Usher and Elliott [15]. An indication that a sample is inhomogeneous can be in addition obtained by comparing the absolute values of theoretical and experimental magnetizations; for strongly inhomogeneous samples the experimental values are much lower than the theoretical ones.

The first results of dHvA effect on 2DEG were obtained using superlattices and they were usually sinusoidal in shape, see e.g. Störmer et al. [52] and Eisenstein et al. [53]. Since the experimental values of magnetization were much lower than the theoretical ones, such data are nowadays interpreted as having been obtained on inhomogeneous samples. More recent investigations were often carried on single QWs which made the interpretations more conclusive. We quote recent results of Wilde et al. [54] given in three parts in Fig. 22, corresponding to GaAs/GaAlAs samples with different spacers. It is seen that in the sample with the widest spacer (lowest part) the 2D electron density $N$ is lowest and the high-field slopes of dHvA oscillations are almost vertical. As the spacers become more narrow (two higher parts) the densities $N$ are higher and the high-field slopes are less and less vertical. This is interpreted by the authors as an indication of a constant background of localized DOS between LLs due to disorder introduced by the donors in the GaAlAs barrier. As mentioned above, the background DOS “slows down” $E_F$ as a function of the field between LLs and results in the nonvertical slopes of magnetization peaks on the high-field sides. According to this interpretation, the disorder is stronger when the spacer is smaller, so DOS between LLs is higher and the nonverticality more pronounced, in agreement with the observations.

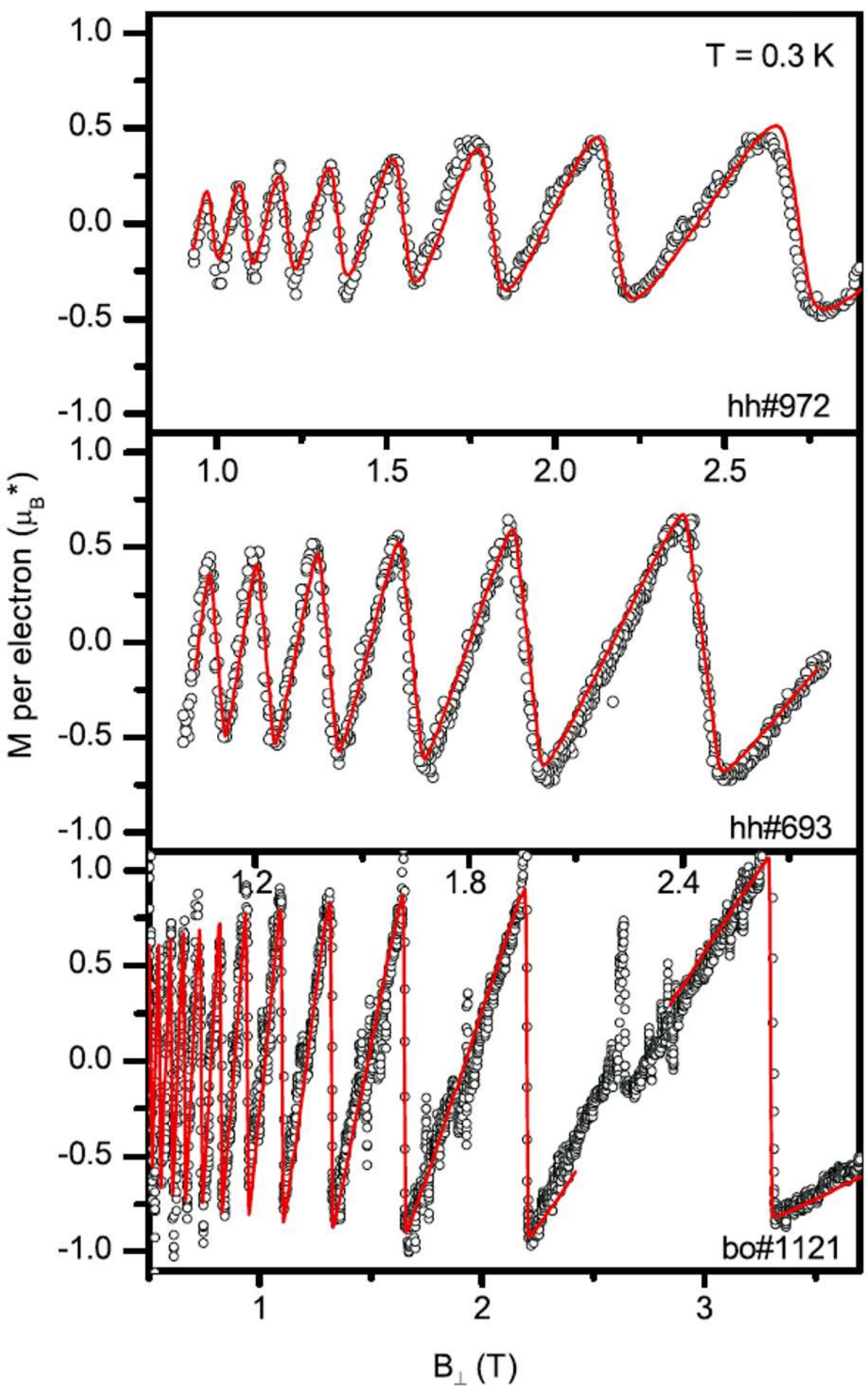

**Figure 22** Magnetization oscillations in GaAs/GaAlAs heterostructures with three different spacers (20, 30, 40 nm) at $T$ = 0.3 K. Empty points – experiments, solid lines – calculations assuming constant DOS between Landau levels. After Ref. [54].

However, one can interpret the same results using the reservoir model. As follows from our Fig. 4, smaller spacers result in higher 2D densities $N$ in the well, which is what one observes, see also Sibari et al. [27]. This point is not controversial. In addition, and this again is seen in Fig. 4, smaller spacers result in stronger transfers of $N$ between the well and reservoir. The transfer “slows down” the Fermi level as a function $B$ between LLs even if there is no background DOS. It is seen in our Fig. 3(a) that in this case the oscillations of $E_F$ have non-vertical slopes on the high-field sides. As a consequence, also magnetization oscillations have non-vertical slopes on the high-field sides since, as follows from our Figs. 2(a) and 2(b) as well as from Figs. 3(a) and 3(c), the behaviour of magnetization follows closely that of $E_F - E_0$. Thus the results shown in Fig. 22 can be equally well explained by the reservoir model. Usher et al. [55] suggested that non-vertical slopes of dHvA magnetization oscillations can result from the electron transfer in the presence of a reservoir.

In conclusions of their work Wilde et al. [54] wrote: “In the highest quality sample (…) we observe a vanishing background DOS and a very small LL broadening. Nevertheless the system shows well defined Hall plateaus in transport measurements. This result sheds new light on theories of the QH effect which relate the QH plateaus width to localized states induced by disorder.” In their recent review Weis and von Klitzing [56] remarked that QHE does not necessarily require disorder to be present. The depletion at the 2D edges and in front of the alloyed contacts might be enough. On the other hand, we can add that the reservoir model explains perfectly well the situation described by Wilde et al. [54].

**7 Magneto-plasmons** An interesting indication of the reservoir model was found in investigations of magneto-plasmons (MP). Holland et al. [16] used photoconductivity spectroscopy to investigate MP in GaAs/GaAlAs QWs with one populated electric subband and found an unusual MP dispersion. To appreciate this result let us briefly consider the MP frequency $\omega_p$. The dispersion of MP in the long-wave limit is [57]

$$\omega_p^2(q) = \frac{2\pi Ne^2}{\varepsilon m^*} q, \quad (20)$$

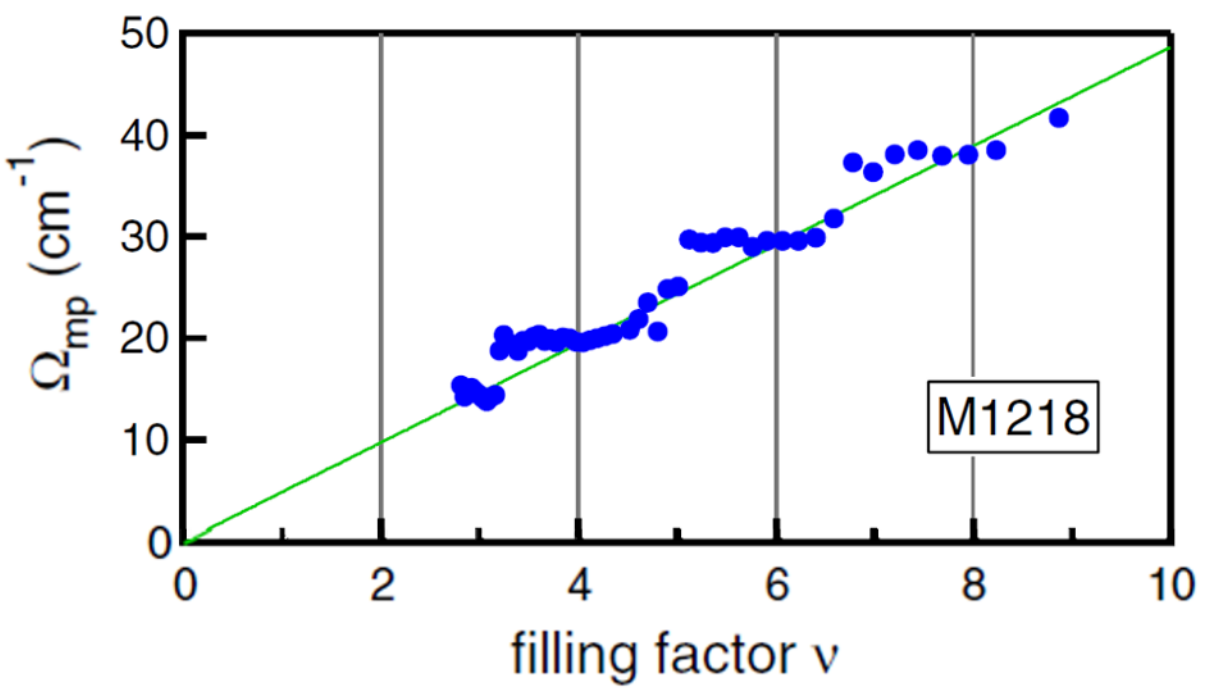

**Figure 23** Filling-factor dependence of renormalized plasma frequency measured on GaAs/GaAlAs sample M1218. The dashed line is a semiclassical prediction. After Ref. [16].

where $\varepsilon$ is the effective dielectric permittivity and $q$ is the plasmon wave vector. In the presence of a transverse magnetic field $B$ in the Voigt configuration the MP frequency is [58]

$$\omega_{mp}^2(q) = \omega_p^2(q) + \omega_c^2, \tag{21}$$

where $\omega_c$ is the cyclotron frequency. One defines a renormalized MP frequency

$$\Omega_{mp} = \frac{\omega_{mp}^2(B) - \omega_c^2}{\omega_c}. \tag{22}$$

By using above equations this can be rewritten as

$$\Omega_{mp} = \frac{2\pi ecN}{\varepsilon B} q = \frac{2\pi e^2}{\varepsilon h} \nu q, \tag{23}$$

where $\nu$ is the filling factor. It follows from Eq. (23) that, for a constant 2D density $N$, the frequency $\Omega_{mp}$ should be a smooth linear function of $\nu$. However, the experimental data shown in Fig. 23 show clearly QHE-like plateaus forming around the even filling factors $\nu$ = 4, 6, 8. These results strongly resemble the quantum Hall plateaus measured in the dc magneto-transport.

Toyoda et al. [17] observed that the results shown in Fig. 23 can be easily explained if one adopts the reservoir hypothesis. Assuming that the Fermi energy is pinned by a reservoir, the electron density $N$ appearing in Eq. (23) oscillates as $B$ increases, see our Fig. 3(b). As a consequence, the ratio $N/B$ exhibits plateaus when $N$ increases linearly with $B$, just like in QHE. This is what one observes. Thus, the result shown in Fig. 23 confirms the reservoir model.

**8 Discussion and conclusions** At the end of our review one can ask a few important questions. The first is: is there enough evidence for the existence of reservoirs in various GaAs/GaAlAs heterostructures? We think that the answer to this question is positive. Although the reservoir model, as proposed by Baraff and Tsui, was concerned with QHE, we think that the evidence for the existence of reservoirs is better provided by optical effects which are easier to interpret. Thus we think that the strongest evidence, as collected above, is given by the cyclotron resonance, photo-magneto-luminescence and magneto-plasmons. It is difficult to believe that the agreement shown in Figs. 16, 17 and 18 for CR in three samples, and in Fig. 21 for MPL in four samples is accidental or that the corresponding interpretations are incorrect. As we mentioned above, the nonlinear B dependence of MPL energies was observed by many authors in numerous GaAs/GaAlAs structures which suggests that the reservoirs are present in almost all samples.

It can be seen above that some magneto-transport data can also be explained by the reservoir model. In principle, the localization model and the reservoir model can peacefully coexist: the fact that the electron density changes with magnetic field does not affect the localization theories of the QHE plateaus. However, seeing how well the reservoir approach describes the Quantum Hall Effect and the corresponding Shubnikov-de Haas effect, one is tempted to explain the quantum magneto-transport solely by the reservoir model. For example, this was done with good results by Raymond and Sibari [10] for GaAs/GaAlAs and recently by Janssen et al. [19] for graphene, see our Section 3. Thus, one is tempted to ask: can one describe the quantum magneto-transport of 2DEG using only the reservoir approach? Here, in our opinion, the answer is negative. First, if one uses the natural reservoir provided by donors in the barrier, as proposed by Baraff and Tsui, the resulting QHE plateaus are usually too narrow. For this reason, in order to get a good description, Raymond and Sibari were forced to assume that the Fermi level was completely pinned. Similarly, Janssen et al. used a very large reservoir. We note in passing that, if the Fermi level were completely pinned by a large reservoir, one would be able to pass the lowest Landau level through the Fermi energy with increasing magnetic field driving all the electrons into the reservoir and the regime of the Fractional QHE would look completely different from what is observed in many experiments.

In most cases of high quality samples the reservoir is not large and, once the electrons fill the reservoir as the field increases, its presence is not felt. Finally, there is the problem of background DOS between LLs. The reservoir model works well if there is no DOS between LLs because only then the filled LLs contain the number of electrons giving the correct Hall quantization. If there exists nonvanishing DOS between LLs, even small (see e.g. Ref. [59]), one will not have the exact quantization when $E_F$ is between LLs. One then needs the localization concepts to guarantee the correct plateaus. Thus we think that in real two-dimensional world one deals with combinations of localization and electron transfer in various proportions. As the temperature is lowered, the role of localization increases.

Intimately related to this problem is the above mentioned validity of classical formula for $\rho_{xy}$. We repeat what we said above: if a quantum Hall plateau is due to the electron transfer, the classical relation $\rho_{xy} = B/Nec$ measures the density $N$ at all fields $B$ and it gives the plateau when $N$ increases linearly with $B$. If, on the other hand, a plateau is due to the localization and $N$ remains constant, the above formula for $\rho_{xy}$ can not be valid since with increasing $B$ and constant $N$ its value is not constant. Thus, the above formula for $\rho_{xy}$ is not valid at plateau fields if the localization is involved. We remark that this formula was implicitly used when interpreting the results for the cyclotron resonance (in the magneto-transport part) and magneto-photo-luminescence, as well as magneto-plasmons, and it is often used explicitly or implicitly in interpretations of magneto-transport, see e.g. Baraff and Tsui [2], Bok and Combescot [3], Xu [24], Raymond and Sibari [10], Janssen et al. [19], etc. The above ambiguity illustrates again our statement that the transport data are difficult to interpret.

An opinion is sometime expressed that reservoir and localization are basically the same thing. The electrons either go to the reservoir outside of the well and cease to conduct (external reservoir) or else go to the localized states in the well and cease to conduct (internal reservoir). We emphasize, however, that the physics giving plateaus of QHE in both cases is completely different. First, the electrostatics of the open structure with an outside reservoir is different from the closed structure. Second, the mechanism of the plateaus in both cases is entirely different. In case of an outside reservoir the mechanism is based on the electron statistics in the presence of a magnetic field. In case of localization, one needs to evoke properties of localized and delocalized 2D states in a magnetic field guaranteeing that the delocalized electrons carry the total current.

We do not consider here the Landauer-Büttiker approach to QHE based on edge states, see Ref. [60]. In their recent review paper based on scanning force microscopy Weis and von Klitzing [56] argued convincingly that in the crucial regions of quantized Hall plateaus the Hall currents flow mostly in bulk of the sample.

Finally, one cannot avoid a natural question concerning the nature of a reservoir. Here the answer will certainly depend on the system under investigation. Baraff and Tsui proposed the natural reservoir for GaAs/GaAlAs structures related to the donors in the barrier introduced by the modulation doping in order to have electrons in the GaAs well. Such a reservoir certainly exists. One clearly has additional donors at the GaAs/GaAlAs interface since an interface provides a natural barrier for diffusing ions. Some authors mention donors at the surface of the whole structure that can pin the Fermi energy. In MOS structures it is usually assumed that there exist donors at the metal-oxide interface. Finally, one should not forget that metallic or alloy contacts used to measure transport effects can also stabilize the Fermi energy in the whole structure. One should clearly look for a specific reservoir in each case under consideration.

**Dedication** We dedicate this work to the memory of our collaborator Stephane Bonifacie, who died suddenly of a heart attack at the age of 32. Stephane was an exceptionally competent and gentle person. He greatly contributed to the subject of this article by doing decisive work on the magneto-photo-luminescence in GaAs/GaAlAs asymmetric quantum wells reported in our Ref. [14]. Stephane's untimely departure has been an irreparable loss to everybody who knew him.